\documentclass[aps,pra,twocolumn]{revtex4}
\usepackage{bm}
\usepackage{amsmath}
\usepackage{amssymb}
\usepackage{graphicx}
\usepackage{epsfig}
\usepackage{epstopdf}
\usepackage{mathrsfs}

\begin{document}

\def\be{\begin{equation}}
\def\en#1{\label{#1}\end{equation}}

\newcommand{\rd}{\mathrm{d}}
\newcommand{\vare}{\varepsilon }
\newcommand{\nb}{\mathbf{n}}
\newcommand{\bs}{\mathbf{s}}
\newcommand{\mb}{\mathbf{m}}
\newcommand{\tb}{\mathbf{t}}
\newcommand{\Phib}{\mathbf{\Phi}}
\newcommand{\Psib}{\mathbf{\Psi}}
\newcommand{\Mc}{\mathcal{M}}
\newcommand{\Zc}{\mathcal{Z}}
\newcommand{\alphb}{{\boldsymbol\alpha}}
\newcommand{\betb}{{\boldsymbol\beta}}
\newcommand{\U}{\mathcal{U}}

\title{Distinguishability in quantum interference with multimode squeezed states}

\author{Valery  Shchesnovich }

%\affiliation{Centro de Ci\^encias Naturais e Humanas, Universidade Federal do
%ABC, Santo Andr\'e,  SP, 09210-170 Brazil }

\begin{abstract}
Distinguishability theory   is developed  for  quantum   interference of  the    squeezed   vacuum states  on  unitary linear  interferometers.    It is found that the entanglement of   photon pairs  over   the Schmidt modes  is one of the sources of  distinguishability.   The distinguishability     is quantified by  the symmetric part of the internal state of $n$ pairs of photons over the spectral Schmidt modes, whose normalization $q_{2n}$ is    the probability   that $2n$ photons interfere as  indistinguishable.   For  two pairs of photons     $q_{4}=(1+ 2\mathbb{P} )/3$, where  $\mathbb{P} $  is  the purity of the squeezed states   ($K=1/\mathbb{P} $ is  the Schmidt number). For  a fixed  purity $\mathbb{P}$, the probability      $q_{2n}$     decreases  exponentially  fast  in  $n$.    For example, in the    experimental  Gaussian boson sampling of   H.-S.~Zhong \textit{et al}   [Science \textbf{370},   1460 (2020)],     the achieved  purity  $\mathbb{P}\approx 0.938$ for the  average number of  photons  $2n\ge 43$  gives  $q_{2n}\lesssim  0.5$, i.e.,  close to the middle line between $n$ indistinguishable and $n$ distinguishable pairs of photons.  In  derivation of  all the results   the first-order quantization representation based on the particle decomposition of the Hilbert space of identical bosons     serves as   an  indispensable  tool.  The approach   can be   applied  also to the   generalized  (non-Gaussian)   squeezed   states, such as those  recently generated in  the three-photon   parametric down-conversion.   

\end{abstract}

%\pacs{ 42.50.St, 03.67.Ac, 42.50.Ar }
\maketitle

\section{Introduction}
\label{sec1}

Non-classical states of light  are  useful  for  the  quantum   information,  computation and interferometry \cite{PhQI1,PhQI2,MultPhInterferom}. The quantum interference  of indistinguishable single photons on   unitary linear multiports can serve as  a basis of the computation  superiority over digital computers,   formulated as the boson sampling  idea  \cite{AA},  a pathway to   demonstration of  the quantum advantage  \cite{20PhBS}. Quantum optical platforms seem to be the most suitable for this  purpose \cite{ReviewBS}.  One can also employ  Gaussian states, instead of  single photons, realizing  the boson sampling  with  Gaussian states    \cite{GBS1,ScattBS,GBS2,ExpGBS1,ExpGBS2}.   Gaussian states  have   been found  useful  in many other quantum information tasks \cite{GaussQI}.   With   Gaussian states one can  efficiently   simulate  quantum chemistry with molecules  \cite{BSmolec,MDGBS} and some computational tasks on graphs  \cite{Graphs1,Graphs2}. These tasks require scaling up the number of Gaussian  states in the  interference experiments,  inciting the search  for scalable sources  \cite{ScalSQS}. 

Scaling up the number of interfering  photons requires strong   control of their  distinguishability due to the fluctuating parameters.   When generalizing the  Hong-Ou-Mandel experiment  \cite{HOM}  to more  than two  photons,  it was  found that the distinguishability is     described by the  symmetric  group    \cite{Ou}.   The effect of  the distinguishability in     interference with identical particles, bosons or fermions, has been  studied in a number of theoretical works  \cite{3phDist,SuffCond,DTh,TBonBS,Tichy,SYMDTh,TL,CollPh,DistTimeRes,QSPD,SWD}  and  experiments   \cite{Dist6Ph,6phbunch,genBB,MultPh1,SYMDTh,DMPI,4ph}.  Signatures of inter-species distinguishability are also revealed in systems of  interacting bosons  \cite{DMBD}.    Distinguishability  degrades the   quantum  advantage  with single photons  \cite{SuffCond,TBonBS},   allowing  for classical simulation of the boson sampling    \cite{SimDistBS}.   Such a dramatic  effect can also be expected  with  the  squeezed  states,  used in  the  experiments on the Gaussian boson sampling     \cite{ExpGBS1,ExpGBS2}.  As the  experimentally obtained squeezed states are multimode (i.e., the purity is not exactly $1$),  one would like to know  if  their multi-mode structure  induces    partial distinguishability. Though some experiments have shown that the multi-mode structure is shaping   the interference  with the squeezed states \cite{ModeStrInfl},    there were no clue of  how one could  approach such a  problem.   
  
The aim  of this work is  to give theory of  distinguishability for  interference of  the squeezed vacuum states on  linear unitary interferometers.   The main result is the output probability distribution,  applicable to the  interference of an  arbitrary number   of  multi-mode squeezed vacuum states on arbitrary linear interferometer.  The  four-photon interference  with the multi-mode  squeezed states    \cite{ModeStrInfl} and the output probability formula  for the single-mode squeezed states    \cite{GBS2}  follow from   the main result   in these special cases.  
The   measure of partial  distinguishability,  analogous to that for single photons \cite{TBonBS}, is found for   interference with the squeezed states.  An estimate  of  the  degree of distinguishability in the   Gaussian boson sampling experiment   \cite{ExpGBS2} follows.      It is illuminating that all the results are easily  derived  by decomposing  the Hilbert space of identical bosons as a direct sum of   tensor powers of the single-particle Hilbert space,   i.e.,  within  the    first quantization   applied to    identical bosons.   The approach  can be applied also to   generalized squeezed states \cite{GenSQ}, such as those    obtained in  the  recently demonstrated three-photon parametric down-conversion \cite{3phDC}.  
 
The rest of the text is organized as follows.  In section \ref{sec2}  it is shown how to rewrite a   squeezed vacuum  state in    the particle decomposition of the Hilbert space of identical bosons, termed here  the first-order  quantization representation. The relation of the latter to  the oscillator decomposition in the usual,  second-order,  quantization, is discussed in subsection \ref{addsec1}.  In subsection \ref{addsec2} the general multi-mode squeezed states are rewritten in this form.  In section \ref{sec3}  the  interference of $N$  squeezed states on a unitary linear interferometer is analyzed.  For the    single-mode  squeezed states  the familiar  expression for  the output  amplitude as  a matrix Hafnian \cite{GBS2}  is recovered in  subsection \ref{sec3A}.   The case of   the multi-mode squeezed states with identical  Schmidt modes  is analyzed in  subsection \ref{sec3B},  where we also recover, as a special case,  the previous results  for the four-photon interference on a beamsplitter \cite{ModeStrInfl}. The case when there are  orthogonal  internal modes  of photon pairs coming from different  sources is   considered in    subsection \ref{sec3C}  and   the  general  case is discussed in  subsection \ref{sec3D}.  In section \ref{sec4} a measure of the   distinguishability in interference with the  squeezed states is proposed and its physical   interpretation is found.  The  partial distinguishability in the recent Gaussian boson sampling experiment of Ref. \cite{ExpGBS2} is characterized there.  Possibility of  application  of the approach  to the generalized (non-Gaussian)   squeezed states is discussed in section \ref{sec5}.   Section \ref{sec6} gives concluding remarks. Some mathematical details,  unnecessary for understanding of the main text,  are placed in Appendices A-E. 

%%%%%%%%%%%%%%%%%%%%%%%%%%%%%%%%%%%%%%%%%%%%%%%% 
\section{Squeezed   states  in the first-order quantization representation  } 
\label{sec2}

Squeezed   states  \cite{ST1,ST2}  are  usually  produced  by the second-order nonlinearity in the process of spontaneous  parametric downconversion \cite{ST3,ST4,ST5}, as well as  the  third-order (Kerr) nonlinearity in the four-wave mixing process  \cite{ST6A,ST6}.   In  the following we will  consider only the squeezed vacuum states, simply referred to as the squeezed states. We will also distinguish between  the  degenerate  and non-degenerate squeezed states, where in a  degenerate squeezed state  photon pairs occupy  the same set of  Schmidt modes (giving the spectral shape),  whereas in a non-degenerate  squeezed state   photon pairs occupy  different Schmidt modes  due to   different   polarizations (and, possibly,  also have different  spectral shapes as well).  

The  squeezed states can be most   conveniently  represented   by the  singular-value decomposition of the  squeezing    Hamiltonian  \cite{MMSS}     (see  also  Refs. \cite{PurityGS,ExpMMSS,BMR}).   In general, there  can be   infinite number of singular values  and the corresponding orthogonal  (a.k.a.  Schmidt) modes. The  degenerate $|r)$ and non-degenerate $|\tilde{r})$  squeezed states   can be always  cast as follows:      
\begin{eqnarray}
\label{E1} 
&& |r) \equiv  Z \exp\Biggl\{ \frac{r}{2}\sum_{j=1}^\infty \sqrt{p_j} \hat{a}_{\phi_j}^{\dag 2}\Biggr\}|0), \\
&& Z = \prod_{j=1}^\infty  \left(1- r^2p_j\right)^{\frac14},  \nonumber\\
&& |\tilde{r}) \equiv  \tilde{Z} \exp\Biggl\{ r\sum_{j=1}^\infty \sqrt{p_j} {\hat{a}_{H,\phi_j}}^\dag  {\hat{a}_{V,\psi_j}}^\dag \Biggr\}|0), 
\label{E2} \\
&& \tilde{Z}  = \prod_{j=1}^\infty  \left(1- r^2p_j\right)^\frac12, \nonumber
\end{eqnarray}    
where   $\hat{a}_{ \phi_j}^\dag$  is the photon creation operator for   Schmidt mode $\phi_j$ and,   similarly,    ${\hat{a}_{H,\phi_j}}^\dag$,  ${\hat{a}_{V,\psi_j}}^\dag $,  with $H,V$ denoting two  orthogonal   polarizations and $j$ generally different spectral modes,  $\phi_j$ for $V$ and $\psi_j$ for $H$,   whereas   $|0)$ denotes   the vacuum state (the tensor  product of the vacuum states in all the modes).   The  polarization is omitted  in the degenerate case for simplicity.    Here and below the notation $| \ldots )$  is  used   for the Fock states and squeezed states,  whereas  the  standard notation $|\ldots\rangle$ is  reserved   for the states in the  first-order quantization representation (see details in section \ref{addsec1}). The   singular-values \mbox{$0\le p_j\le 1$} are  conveniently  normalized, where  $\sum_{j=1}^\infty  p_j = 1$ (this choice will become clear below), whereas the  normalization  factor $0<r< 1$  (we   consider $r$ to be real as  the possible phase  factor  can be incorporated into the boson creation operators) will be called     the   squeezing parameter  ($r = \tanh\kappa$, where $\kappa$ is usually called the squeezing parameter).    For a   single-mode squeezed state ($p_1=1$)  the parameter $r$ in Eqs. (\ref{E1})-(\ref{E2}) is related to   the average number of detected photons $2\bar{n}$ as follows    $2\bar{n} = r^2/(1-r^2)$ (i.e., $2\bar{n} =  \sinh^2\kappa$). 
The  set of singular-values $\{p_1,p_2,\ldots\}$   characterizes  multimodeness  of the squeezed  state,   which can be   quantified either by the purity $0< \mathbb{P}\le 1$ or by  the Schmidt number $1\le K <\infty$ \cite{MMSS,PurityGS,ExpMMSS,BMR,ModeStrInfl}, where   
\be
\mathbb{P}  = \sum_{j=1}^\infty  p_j^2,\quad K = \frac{1}{ \mathbb{P}}. 
\en{SchmidtK}
The  Schmidt number was  recently  shown to shape the  four-photon interference \cite{ModeStrInfl}. 
 
 In  Eqs. (\ref{E1})-(\ref{E2}) we have tacitly assumed that  the  squeezed  states   are pure states, i.e.,  that there is    perfect cross-photon-number coherence, which is sometimes  argued to be unnecessary for understanding the experiments     \cite{PhNumIn,noSS}.  The squeezed states   in  Eqs. (\ref{E1})-(\ref{E2}) represent the usual  parametric approximation,  applicable  when  the   pump   is sufficiently strong and  the  interaction times are sufficiently short    \cite{PumpEff}.   This  approximation   disregards  the precise balance of annihilated and created photons  due to the   energy conservation \cite{SPent,EnConDC}. Taking such a balance  into account would result in   imperfect coherence between the multi-photon components with different $n$, due to  entanglement with  generally  different quantum  states of the pump.    Here,  we   disregard such effects, relegating  their study to future publications. 
 
 Below, we will  derive another, more useful for our purposes,  representation of  the  squeezed states.  Our representation  utilizes another possible decomposition  of the Hilbert space of identical bosons, in contrast to  the standard decomposition by independent oscillators. The relation between  the two is discussed   below. 
 
%%%%%%%%%%%%%%%%%%%%%%%%%%%%%%%%%%%%%%%%%%%%
 \subsection{ The  first-order and second-order quantization representations  for identical bosons } 
\label{addsec1}

The  quantization of the electromagnetic  field, historically termed    the second quantization,   is usually performed    by  representing it as a system of independent oscillators  in some orthogonal modes.  In this approach   the  Hilbert space of quantum states of photons   is decomposed as  the  tensor product of the  Hilbert spaces of independent oscillators.

The term first quantization  refers to    quantum description of    particles, such as a system of   identical bosons.     In this approach the  symmetric subspace of the tensor power  of the Hilbert space  of individual bosons is the physical Hilbert space  (in general,  it is  the direct sum of such subspaces,  when the number of bosons is not fixed).  

Photons are  bosons, hence  there is a mathematical equivalence  between the above  two approaches.   Below this  mathematical equivalence  is exposed   following   Ref.  \cite{MyNotes}   (similar approach is used in Quantum Statistical Mechanics by Bogolubov \cite{BogBog} and the  essential  features are found already in Dirac \cite{Dirac}).   We will use the terms  ``second-order  quantization" and `first-order quantization"  referring to the  above two decompositions of the Hilbert space of identical bosons. 

 In the first-order quantization approach     $n$ identical bosons  occupy a completely symmetric state  in  the  tensor  power space  $\mathcal{H}^{\otimes n}$,  where  $\mathcal{H}$  is the Hilbert space of  a single boson   (i.e., of the single-particle states).    If the occupied single-particle    states are  some  orthogonal states $|\psi_k\rangle \in \mathcal{H}$, $k=1,2,\ldots$, with occupations, say,  $\nb = (n_1,n_2,\ldots,n_m)$,   $|\nb|= n_1+n_2+\ldots+n_m= n$ (and the rest $n_j =0$),  then the   state of $n$ bosons in $\mathcal{H}^{\otimes n}$ in $m$ modes is the following  Fock state
\begin{eqnarray}
\label{S1}
&&|n_1,n_2,\ldots,n_m)^{(I)} \equiv   \sqrt{\frac{n!}{\nb!}} \hat{S}_n|\psi_{k_1}\rangle\ldots|\psi_{k_n}\rangle\nonumber\\
&& \qquad= \sqrt{\frac{n!}{\nb!}} \hat{S}_n|\psi_1\rangle^{\otimes n_1} \dots |\psi_m\rangle^{\otimes n_m},
\end{eqnarray}
where $k_1,k_2,\ldots, k_n$ is the multi-set of indices corresponding to the non-zero occupations $\nb$, $\nb!=n_1!n_2!\ldots n_m!$,     and  $\hat{S}_n$ is the projector on the symmetric subspace of $\mathcal{H}^{\otimes n}$ defined as follows:
\begin{eqnarray}
\label{Sn}
&& \hat{S}_n = \frac{1}{n!} \sum_{\sigma\in S_n} \hat{P}_\sigma, \\
&&   \hat{P}_\sigma  |\phi_1\rangle |\phi_2\rangle\ldots |\phi_n\rangle  \equiv  |\phi_{\sigma^{-1}(1)}\rangle |\phi_{\sigma^{-1}(2)}\rangle\ldots |\phi_{\sigma^{-1}(n)}\rangle \nonumber,
\end{eqnarray}
where $\sigma$ is  an element (permutation) of  the symmetric group $S_n$ of $n$ objects. Due to the group property,  $\hat{P}_\sigma \hat{P}_\tau = \hat{P}_{\sigma\tau}$, we have   $\hat{P}_\sigma\hat{S}_{n}  =  \hat{S}_n $ and $\hat{S}^2_n = \hat{S}_n$, thus $\hat{S}_n$  is a projector on the symmetric subspace of the  tensor  power of  $\mathcal{H}^{\otimes n}$, denoted below $\hat{S}_n\left\{\mathcal{H}^{\otimes n}\right\}$.  More generally, when  $n$ bosons occupy some   arbitrary single-particle  states $|f_1\rangle,\ldots,|f_n\rangle\in \mathcal{H}$, then the following     unnormalized  state 
\be
|f_1,\ldots,f_n\rangle \equiv   \hat{S}_n|f_1\rangle\ldots|f_n\rangle
\en{S2}
corresponds to this case.  The normalization factor for the state in Eq. (\ref{S2})  can be derived  from  the inner product of  two symmetric states, which reads 
\be
 \langle g_1,\ldots,g_n |f_1,\ldots,f_n\rangle  = \frac{1}{n!}\sum_\sigma \prod_{i=1}^n \langle g_i|f_{\sigma(i)}\rangle.
\en{S3}
Any state of $n$ bosons is also    some   linear combination of Fock states, as in  Eq. (\ref{S1}),   in any given basis   in $\mathcal{H}$. Thus,  the state in Eq. (\ref{S2}) can be rewritten as such by expansion of the single-particle states $|f_k\rangle$ in a given   basis.

The   equivalence between the first-order and second-order quantization  of identical bosons can be established by  introducing the equivalents of   the  boson creation and annihilation   operators   as  some  linear operators acting   between the symmetric subspaces  $\hat{S}_n\left\{\mathcal{H}^{\otimes n}\right\}$ with different $n$. Consider the following   two operators  \cite{MyNotes}:
\begin{eqnarray}
&&  \hat{A}^+_\phi | f_1,\ldots,f_n\rangle \equiv   \sqrt{n+1}\hat{S}_{n+1}|\phi\rangle |f_1,\ldots,f_n\rangle, \nonumber\\
&&  \hat{A}^-_\phi |f_1,\ldots,f_n\rangle \equiv   \sqrt{n} (\langle\phi|\otimes I\otimes\ldots\otimes I) |f_1,\ldots,f_n\rangle\nonumber\\
&& \quad= \frac{1}{\sqrt{n}}\sum_{i=1}^n\langle\phi|f_i\rangle |f_1,\ldots,f_{i-1},f_{i+1},\ldots, f_n\rangle,
\label{S4}
 \end{eqnarray}
 where for $\hat{A}^-_\phi $ we have used the expansion $S_{n} = \sum_{i=1}^n (1,i)S^{(i)}_{n-1}$, with $(1,i)$ being the transposition of $1$ and $i$ (fixed point for $i=1$) and $S^{(i)}_{n-1}$ the symmetric group of permutations of $(1,2,\ldots,i-1,i+1,\ldots, n)$. 
By definition, the operator $\hat{A}^+_\phi$ acts by adding a  boson in the state $|\phi\rangle$ to the state it applies to (and symmetrization), whereas,   the operator $\hat{A}^-_\phi$ acts by  removing  a  boson in the state $|\phi\rangle$ (replacing a single-particle state by the amplitude of its  projection  on $|\phi\rangle$).  One can show  \cite{MyNotes} that the introduced operators satisfy the following properties:
\begin{eqnarray}
&&(\hat{A}^{\pm}_\phi)^\dag = \hat{A}^{\mp}_\phi, \nonumber\\
&& [\hat{A}^{\pm}_\psi, \hat{A}^{\pm}_\phi ] = 0, \nonumber\\
&& [\hat{A}^{-}_\psi, \hat{A}^{+}_\phi ] =
\langle\psi|\phi\rangle,
\label{S5}
\end{eqnarray}
where $[\hat{A},\hat{B}] = \hat{A}\hat{B}-\hat{B}\hat{A}$ is the commutator. For  a basis $|\psi_k\rangle$, $k=1,2,\ldots,$   in $\mathcal{H}$, we recover the usual commutation relations for the boson creation and annihilation operators by  associating   
\be
\hat{a}^\dag_{\psi_k} = \hat{A}^+_{\psi_k}, \quad  \hat{a}_{\psi_k} = \hat{A}^-_{\psi_k}.
\en{S6}

 The final step   is to  complete the sequence of the  tensor powers  $\mathcal{H}^{\otimes n}$  to that with  $n\ge 0$ by adding   the $0$-power  $\mathcal{H}^{\otimes 0}\equiv \{|\;\rangle\}$ (i.e., the state containing no particles; see also Ref. \cite{Dirac}) by postulating 
  $|\;\rangle \equiv |0) $, where $|0)\equiv \prod_k|0)_k$ is the  tensor product of the  individual  vacuum states  $|0)_k$  of all   modes in some basis. Then, a repeated application of the definition of the creation operator in Eq. (\ref{S4})  to the vacuum state   gives 
  \be 
  \hat{a}^\dag_{\phi_1}\ldots \hat{a}^\dag_{\phi_n} |0)  = \sqrt{n!}\hat{S}_n   |\phi_1\rangle |\phi_2\rangle\ldots |\phi_n\rangle, 
 \en{E3}
where   the boson  operators $\hat{a}^\dag_{\phi_1},\ldots, \hat{a}^\dag_{\phi_n}$  create  arbitrary (i.e., non-orthogonal, in general)  single-particle states  $|\phi_1\rangle, \ldots, |\phi_n\rangle \in \mathcal{H}$.      For example, Eq. (\ref{E3}) relates the  Fock state of $n$ bosons in $m$ modes in the second-order quantization representation and its equivalent Fock state in the first-order quantization representation, Eq. (\ref{S1}), \cite{Dirac}:
\begin{eqnarray}
\label{Fock} 
&& |n_1,n_2,\ldots,n_m)^{(II)}\equiv \frac{(\hat{a}^\dag_{\psi_1})^{n_1}\ldots (\hat{a}^\dag_{\psi_m})^{n_m}}{ \sqrt{\nb!} }|0)  \\
&& =  \sqrt{\frac{n!}{\nb!}} \hat{S}_n  
|\psi_1\rangle^{\otimes n_1} \dots |\psi_m\rangle^{\otimes n_m}= |n_1,n_2,\ldots,n_m)^{(I)},\nonumber
\end{eqnarray} 
  where     $\nb=(n_1,n_2,\ldots, n_m) $, $|\nb|\equiv n_1+n_2+\ldots + n_m=n$.  The  oscillator modes  themselves  form   a basis of the Hilbert space of single-partile states $ \mathcal{H}$. 
  
In summary, the Hilbert space of identical bosons $\mathscr{H}$ in the second-order quantization representation  is  the  tensor product of  the Hilbert spaces  $H_k$ of the orthogonal oscillators in a basis of modes $|\psi_k\rangle\in \mathcal{H}$, whereas in the first-order quantization representation it is a direct sum of the symmetric subspaces of the tensor powers  of the Hilbert space $\mathcal{H}$  of individual bosons:
\be
\mathscr{H} = \prod_{k=1}^{\mathrm{dim}  \mathcal{H}}\!{}^{\otimes} H_k  = \sum_{n=0}^\infty{\!}^\oplus \hat{S}_n\left\{\mathcal{H}^{\otimes n}\right\},
\en{Hilbert}  
 where   $\hat{S}_0 = \openone$. 
The action of the boson creation and annihilation operators for a given   basis of modes can be represented schematically as follows:
\begin{eqnarray} 
\label{Opers}
 && H_k \xrightarrow{\hat{a}^\dag_{\psi_k},\hat{a}^{}_{\psi_k}} H_k,\nonumber\\
 && \hat{S}_n\left\{\mathcal{H}^{\otimes n}\right\} \xrightarrow{\hat{a}^\dag_{\psi_k}}\hat{S}_{n+1}\left\{\mathcal{H}^{\otimes (n+1)}\right\},\nonumber\\
 && \hat{S}_n\left\{\mathcal{H}^{\otimes n}\right\} \xrightarrow{\hat{a}^{}_{\psi_k}}\hat{S}_{n-1}\left\{\mathcal{H}^{\otimes (n-1)}\right\},
\end{eqnarray} 
where in the last line $n\ge 1$, whereas  for  $n=0$ we have  $ \hat{S}_0\left\{\mathcal{H}^{\otimes 0}\right\}   \xrightarrow{\hat{a}^{}_{\psi_k}} 0$. 
  
Finally,  below we will also utilize  a factorization of the oscillator modes, such as the Schmidt modes in Eq. (\ref{E1})-(\ref{E2}), into  the spatial mode,  below denoted   simply by $|k\rangle$, $k=1,\ldots, M$, and corresponding to input port of an unitary interferometer   the squeezed state is launched to (this nomenclature will include also  the polarization, fixed in the degenerate case, but  also in the non-degenerate case, see section \ref{sec3})  and the internal states $|\phi^{(k)}_j\rangle$, $j\ge 1$. In this  case the single-particle state becomes  the product of such states (corresponding  to different degrees of freedom of a photon), e.g., in the case of the degenerate squeezed state launched into input port $k$ of the interferometer, the single-particle states become   $|k\rangle|\phi^{(k)}_j\rangle\in \mathcal{H}$. Thus,  two-index  (sometimes three-index)  notation for the creation and annihilation operators will be used (the third index giving the polarization).

 \subsubsection{Linearity of the boson operators $\hat{A}^\pm$ }
  
 Note that by the definition the  Fock space operator $\hat{A}^+_\phi$ is  linear in the state it  ``creates": 
 \be
 |\phi\rangle =  a |f\rangle + b |g\rangle\quad  \Rightarrow \quad  \hat{A}^+_\phi = a\hat{A}^+_f + b\hat{A}^+_g.
\en{S7}
 The   importance of this property  in  linear optics can be now appreciated. If the Hilbert space  $\mathcal{H}$ has a finite dimension $M$, e.g., as for an interferometer, the change of the basis $|\psi_k\rangle \to |\phi_k\rangle$, $k=1,\ldots, M$,   by some unitary matrix $U$, 
\be
|\psi_k\rangle = \sum_{k=1}^M U_{kl}|\phi_l\rangle, 
\en{S8}
induces the respective transformation for the  operators.  Indeed, by the linearity  property in Eq. (\ref{S7}), the corresponding transformation of the creation operators is 
\be
\hat{a}^\dag_{\psi_k} = \sum_{k=1}^M U_{kl} \hat{a}^\dag_{\phi_l}, \quad \hat{a}^\dag_{\phi_l} \equiv \hat{A}^+_{\phi_l}.
\en{S9}
The linear transformation in Eq. (\ref{S8})  is the result of the    unitary  evolution   
\be
\hat{a}^\dag_{\phi_k} =\hat{U}^\dag\hat{a}^\dag_{\psi_k} \hat{U},\quad \hat{U} \equiv \exp\left\{i\sum_{k,l=1}^M \hat{a}^\dag_{\psi_k}E_{lk}\hat{a}^{}_{\psi_l} \right\}, 
\en{S10}
where the Hermitian matrix $E$ is obtained from the exponent of the matrix $U$: $U = e^{iE}$. Observe that in Eq. (\ref{S8}) and its consequence (\ref{S9}) the words ``linear  optical operation"  have clear physical  interpretation:   the single-particle states (the states in  $\mathcal{H}$) for each boson in a multi-boson state are simply expanded in another basis.

One   comment  on   the usage of Eq. (\ref{E3})   is in order.  The projector  $\hat{S}_n$   on the r.h.s. of Eq. (\ref{E3})   depends on the \textit{total} number of bosons,  i.e.,   the factorization by mode property of the oscillator decomposition of the Hilbert space has no equivalent in the particle decomposition.        
For example,  Eq. (\ref{E3})  cannot be used to get an  equivalent first-order quantization representation of a single-mode  Fock state $|n)_k\equiv \left(\hat{a}^\dag_{\psi_k}\right)^n    |0)_k$, where  $|0)_k\in H_k$  but   $|0)_k\notin \mathscr{H} $ (when there are other oscillator modes). The common vacuum $|0)=\prod_k|0)_k \in \mathscr{H} $  should be used in Eq. (\ref{E3}), which relates the above  two  decompositions of the  whole Hilbert space $\mathscr{H} $ of identical bosons.  Note that  it  does not matter   which orthogonal basis of modes  in the Hilbert space $ \mathcal{H}$ is used for  the  mode  factorization  of the common vacuum state $|0)$, since any linear  evolution, Eqs. (\ref{S9})-(\ref{S10}),  leaves   it  invariant. 
   
 The first-order quantization representation  can be used to simplify calculations of the quantum probabilities of photon detection at output of   a linear interferometer,  since it allows the  decomposition the  photon  degrees of freedom  into two classes \cite{SuffCond,DTh}: the operating modes  affected by  interferometer according to  Eqs. (\ref{S8})-(\ref{S9}) and the internal states (or modes) which are invariant under the action of the interferometer.

 %%%%%%%%%%%%%%%%%%%%%%%%%%%%%%%%%%%%%%
\subsection{Squeezed vacuum states in the first-order quantization representation}
\label{addsec2}

Let  us find   the first-order quantization representation of the squeezed  states of  Eqs. (\ref{E1})-(\ref{E2}).  Consider first the degenerate case. It can be viewed as the    tensor product of the single-mode squeezed states,  indexed by $j$  in Eq. (\ref{E1}),     and having  the  squeezing parameters   $r_j = r\sqrt{p_j}$.   Introduce  $\nb = (n_1,n_2,\ldots)$, where $2\nb$ is   the      vector of Fock occupation numbers of the  Schmidt   modes.    We will use the following identity between   the summation over   the occupations  $\nb$  and the   product of independent summations over    the Schmidt modes, 
 \be
 \sum_{|\nb|=n}f(\nb) =   \sum_{j_1=1}^\infty \ldots \sum_{j_n=1}^\infty \frac{\prod\limits_{j=1}^\infty n_{j }! }{n!} f(\nb),
 \en{E5}
 valid for an arbitrary  function  $f(\nb)$  of  the occupations.  Using  the identity of Eq. (\ref{E5})  we obtain   
\begin{eqnarray}
 \label{E4}
&& |r) = \prod_{j=1}^\infty  |r_j) =  Z\sum_{\nb} \prod_{j=1}^\infty \left(\frac{r_j}{2}\right)^{n_j} \frac{ (\hat{a}_{\phi_j}^\dag  )^{2n_j}}{n_j!} |0) \nonumber\\
&& =  Z \sum_{n=0}^\infty \frac{1}{n!} \sum_{j_1=1}^\infty \ldots \sum_{j_n=1}^\infty  \prod_{\alpha=1}^n  \frac{r_{j_\alpha}}{2}  \hat{a}_{\phi_{j_\alpha}}^{\dag 2}    |0)  \nonumber\\
&& =  Z \sum_{n=0}^\infty \binom{2n}{n}^{\frac12} \sum_{j_1=1}^\infty \ldots \sum_{j_n=1}^\infty   \prod_{\alpha=1}^n  \frac{r_{j_\alpha}}{2}\nonumber\\
&& \times\hat{S}_{2n} |\phi_{j_1}\rangle |\phi_{j_1}\rangle \ldots |\phi_{j_n}\rangle |\phi_{j_n}\rangle \nonumber\\
&& =   Z \sum_{n=0}^\infty \binom{2n}{n}^{\frac12} \hat{S}_{2n} \Biggl(\frac{r}{2}\sum_{j=1}^\infty\sqrt{p_j} |\phi_j\rangle |\phi_j\rangle  \Biggr)^{\!\!\otimes n},
\end{eqnarray}  
 where  we have introduced    the  Schmidt mode $\phi_j$, such that $|\phi_j\rangle= \hat{a}^\dag_{\phi_j}|0)$,   and used  Eq. (\ref{E3})    for the first-order quantization representation of  the $2n$-photon state,
 \be
  \prod_{\alpha=1}^n \hat{a}_{\phi_{j_\alpha}}^{\dag 2}    |0) = \sqrt{(2n)!} \hat{S}_{2n}   \prod_{\alpha=1}^n|\phi_{j_\alpha}\rangle^{\otimes 2}.
 \en{Eq_add}
 Quite similarly, we get the first-order quantization representation of the non-degenerate squeezed vacuum state
\begin{eqnarray}
 \label{E6}
&& |\tilde{r}) =\prod_{j=1}^\infty |r_j) =  {Z}\sum_{\nb} \prod_{j=1}^\infty{r_j}^{n_j} \frac{\left(\hat{a}_{H,\phi_j}^\dag \hat{a}_{V,\psi_j}^\dag\right)^{n_j}}{n_j!} |0) \nonumber\\
&& =    {\tilde{Z}} \sum_{n=0}^\infty \binom{2n}{n}^{\frac12} \hat{S}_{2n} \Biggl(r\sum_{j=1}^\infty\sqrt{p_j}  |H,\phi_j\rangle |V,\psi_j\rangle  \Biggr)^{\!\!\otimes n},\nonumber\\
\end{eqnarray}  
where    $|H,\phi_j\rangle = |H\rangle|\phi_j\rangle= \hat{a}_{H,\phi_j}^\dag |0)$ and $|V,\psi_j\rangle =  |V\rangle|\psi_j\rangle=\hat{a}_{V,\psi_j}^\dag |0)$.  

Above we have used a simplified nomenclature, omitting the spatial mode index (indicating the  input port of an interferometer where  the squeezed state is launched), since we have considered  a single such  multimode squeezed state.  Below, therefore, we will have to add another index to the boson operators, indicating the input port  where  the state is launched, e.g.,  $|k\rangle|\phi_j\rangle= \hat{a}^\dag_{k,\phi_j}|0)$ and  $|k\rangle|H\rangle|\phi_j\rangle= \hat{a}^\dag_{k,H,\phi_j}|0)$, etc, where we have denoted by index $k$ the spatial mode  corresponding to  interferometer input port $k$.  To avoid the confusion with the input port operators, will  use the notation $\hat{b}^\dag_{l,\phi_j}|0)=|l^{(out)}\rangle|\phi_j\rangle$, etc,  for the boson creation operators in the spatial output  port $l$ of an interferometer.

One can convert  a  non-degenerate    single-mode  squeezed state   into two degenerate single-mode squeezed states in two different spatial modes by a unitary transformation.   Such a transformation can be physically realized by first separating the polarizations into two spatial modes by the  polarizing beamsplitter, changing one of the polarizations by a wave plate  and using  a balanced beamsplitter on the output modes. The   effect   on the   two  boson operators, say $ \hat{a}_{1,H}^\dag$ and $ \hat{a}_{1,V}^\dag$,  in the non-degenerate squeezed state  of  the same  spatial mode   can be accounted for by the  following   unitary  transformation 
\be
\left( \begin{array}{c} \hat{a}_{1,H}^\dag \\  \hat{a}_{1,V}^\dag\end{array} \right) = \left(\begin{array}{cc}\frac{1}{\sqrt{2} }& \frac{i}{\sqrt{2}}\\
\frac{1}{\sqrt{2} } & \frac{-i}{\sqrt{2}}  \end{array}\right)\left( \begin{array}{c} \hat{b}_{1,V}^\dag \\  \hat{b}_{1,V}^\dag\end{array}\right),
\en{PBS}
which will be called below the polarization-to-propagation mode beamsplitter, where $\hat{b}_{1,V}^\dag, \hat{b}_{2,V}^\dag$ describe  two  output  spatial modes  of  the same polarization. The  product of  two boson operators in the exponent of a non-degenerate squeezed state transforms as follows   
\be
\hat{a}_{1,H }^\dag \hat{a}_{1,V}^\dag = \frac12\left( \hat{b}_{1,V}^{\dag2} + \hat{b}_{2,V}^{\dag2}\right).
\en{Effect}

 We have seen above that a  two-mode squeezed state  in two different polarizations  becomes a product of two single-mode
squeezed states in another basis (which requires  using an interferometer).  A two-mode squeezed
state can always be represented as a product of two single-mode squeezed states \cite{Milburn}.   This is a
particular case of the so-called Bloch-Messiah reduction \cite{BMRed}: a Gaussian unitary acting on the vacuum state  can always be represented as a product of single-mode squeezers in a properly chosen modal basis.   The physical  setup for such a mathematical  transformation  necessitates  using   an  interferometer.  The multi-mode squeezed states in Eqs. (\ref{E1})-(\ref{E2}) are  tensor products  of the single-mode states over the Schmidt modes.  However, the spectral   Schmidt modes are not affected by spatial interferometers \cite{MMSS,PurityGS,ExpMMSS,ModeStrInfl,BMR}.  Therefore,    despite the existence of a formal mathematical equivalence,   the   multimode  squeezed states at input of a spatial interferometer  are  not   equivalent to   the single-mode squeezed states  at input of any other   spatial interferometer.  The multimode structure of such squeezed states  affects the interference of them  on spatial interferometers, as  demonstrated already  in Ref. \cite{ModeStrInfl}. Moreover, as we will see below,  the spectral Schmidt modes affect interference of the squeezed states in a similar way as the mixed internal  states  of single photons do.  Therefore, in accordance with terminology   used for single photons  \cite{3phDist,SuffCond,DTh,TBonBS,Tichy,SYMDTh,TL,CollPh,DistTimeRes,QSPD,SWD}, the  spectral  Schmidt modes will be called below   the \textit{internal modes}.    In   Eqs.  (\ref{E4}) and (\ref{E6})  the   internal     modes are the states $|\phi_j\rangle$ and $|\psi_j\rangle$.   Observe also that   degenerate and non-degenerate squeezed states   are not always  equivalent in  quantum interference experiments.    The non-degenerate squeezed state  of  Eq. (\ref{E6}) may have  different internal modes for different polarizations  ($|\phi\rangle\ne |\psi\rangle$), unlike the degenerate squeezed state of Eq. (\ref{E4}).  Such  states cannot  be transformed into one another by a  spatial interferometer, since the  unitary transformation   \cite{Milburn,BMRed}  relating them  has to act  on   the internal modes    (e.g.,   
in this case we would have the product   $ \hat{a}_{1,H,\phi}^\dag \hat{a}_{1,V,\psi}^\dag $ on the l.h.s. of  Eq. (\ref{Effect}), with the internal modes $|\phi\rangle \ne |\psi\rangle$,  and no linear unitary  transformation not affecting the internal modes would transform  such a product into the sum of  two  operators squared).

%%%%%%%%%%%%%%%%%%%%%%%%%%%%%%%%%%%%%%%%%%%
\section{Output probability from interference of  squeezed states    } 
\label{sec3}

\begin{figure}[h]
\begin{center}
      \includegraphics[width=.325\textwidth,height=0.32\textwidth]{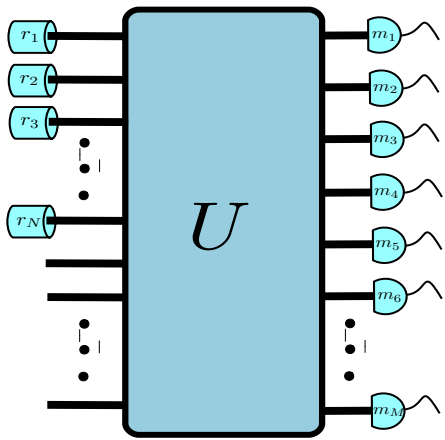} 
     \caption{A schematic depiction of  the considered  setup. Here  $N$ degenerate squeezed vacuum states with squeezing parameters $r_1,\ldots, r_N$  are  launched at the input  of a linear  $M$-port interferometer given by  a unitary matrix $U$. At the output   $M$  particle  number resolving detectors give a configuration $m_1,\ldots,m_M$, with $m_l$  particles detected at output port $l$.   In the non-degenerate case, the two  polarization modes from squeezed source $k$ are assumed to be  launched into inputs $k$ and $N+k$.       \label{F1} }
   \end{center}
\end{figure}

We will consider   quantum interference of  $N$ multimode squeezed vacuum states  having the   overall squeezing parameters $r_1,\ldots, r_N$,  and   impinging on an $M$-port interferometer,  represented here by a unitary matrix $U$ (schematically depicted in Fig. \ref{F1}).     In the non-degenerate case we assume that  photons of different polarizations  are allowed to   interfere     (e.g., by using the polarizing  beam splitters and the  wave plates as components of the interferometer $U$) and, without loss of generality,   that    photons  from source $k$ of the two different polarizations are launched into  input ports $k$ and $N+k$.  We will use the simplified nomenclature incorporating the polarization of photons into the port number,  say  that  the $H$-polarized photons      are launched to ports $1,\ldots,N$, while the $V$-polarized  ones to the ports $N+1, \ldots, 2N$ (the single-index nomenclature    will be  also used for   the  output ports of the  interferometer).   Below, we will  use index $k$ exclusively for the input ports, index  $l$ for the output ports. We are interested in the probability to detect $2n$ photons in an output configuration   $\mb = (m_1,\ldots,m_M)$, where $m_l$ photons are detected at output port $l$, see Fig. \ref{F1}.     Here   $0\le n<\infty$ and  can be    arbitrary,  not  related to $N$ or $M$. Our    model  is applicable   also  to   the  experimental  realization of  Gaussian boson sampling   \cite{ExpGBS1,ExpGBS2}.

The  most general multi-mode squeezed state  at   input port $k$  in the degenerate case  and at  input ports $k$ and $N+k$ in the non-degenerate case are   as follows:
\begin{eqnarray}
\label{Squeez1} 
&& |r_k) \equiv  Z_k \exp\Biggl\{ \frac{r_k}{2}\sum_{j=1}^\infty \sqrt{p^{(k)}_j} \hat{a}_{k,\phi^{(k)}_j}^{\dag 2}\Biggr\}|0), \\
&& Z_k = \prod_{j=1}^\infty  \left(1- r_k^2p^{(k)}_j\right)^{\frac14},  \nonumber\\
&& |\tilde{r}_k) \equiv  \tilde{Z_k} \exp\Biggl\{ r_k\sum_{j=1}^\infty \sqrt{p^{(k)}_j} {\hat{a}_{k,\phi^{(k)}_j}}^\dag  {\hat{a}_{N+k,\psi^{(k)}_j}}^\dag \Biggr\}|0), \qquad
\label{Squeez2} \\
&& \tilde{Z_k}  = \prod_{j=1}^\infty  \left(1- r_k^2p^{(k)}_j\right)^\frac12, \nonumber
\end{eqnarray}    
where $k=1,\ldots, N$ and for brevity we use the same notations for the creation operators in the degenerate and non-degenerate cases (where in the latter case the photon polarization is incorporated into the input  port index).

Interference of  single photons on a linear unitary interferometer is usually  analyzed  by   splitting  the degrees of freedom of photons into   operating modes, acted upon by the interferometer,  and  internal modes, unaffected by the interferometer  \cite{HOM,Ou,SuffCond,DTh,Tichy,SYMDTh,TL,CollPh,DistTimeRes,QSPD,SWD}.  Here,  we have  \mbox{$\hat{a}^\dag_{k,\phi^{(k)}_j}|0) = |k\rangle|\phi^{(k)}_j\rangle$,} where on the right-hand side we   split the single-particle state of a photon into  the operating mode ($|k\rangle$),  acted upon by the  interferometer $U$,   and   the internal state  ($|\phi^{(k)}_j\rangle$),  unchanged by the interferometer.  

A unitary linear interferometer, see Fig. \ref{F1},   in the  first-order  quantization representation  expands   the  basis of  input   modes $|k\rangle$, $k=1,\ldots, M$,   over  the output      basis $|l^{(out)}\rangle$, $l=1,\ldots, M$, where the unitary matrix $U$  gives  the  expansion.     
 From section \ref{addsec1}  we can also get the relation between the boson operators:
 \be
 |k\rangle = \sum_{l=1}^M U_{kl} |l^{(out)}\rangle\quad \Rightarrow  \quad  \hat{a}^\dag_{k,\phi} =  \sum_{l=1}^M U_{kl} \hat{b}^\dag_{l,\phi},
\en{E10}
where  $\hat{b}^\dag_{l,\phi} |0) =  |l^{(out)}\rangle |\phi\rangle$ and it is assumed that the interferometer does not affect the internal states.

Below we will use the  internal state of a photon pair.  The internal state of a photon pair    coming from the input port  $k$ in  the degenerate case will be denoted by $|\Phib^{(2)}_{k}\rangle$  and that in the non-degenerate by $|\widetilde{\Phib}^{(2)}_{k}\rangle$, where 
\begin{eqnarray}
 \label{E7} 
  && |\Phib^{(2)}_{k}\rangle \equiv \sum_{j=1}^\infty\sqrt{p^{(k)}_j}|\phi^{(k)}_j \rangle|\phi^{(k)}_j\rangle,  \nonumber\\
  &&    |\widetilde{\Phib}^{(2)}_{k}\rangle \equiv \sum_{j=1}^\infty \sqrt{p^{(k)}_j}|\phi^{(k)}_j \rangle|\psi^{(k)}_j\rangle.
  \end{eqnarray}
We will see below (in subsection \ref{sec3B} and in section \ref{sec4})    that even when   the squeezed states at different input ports have   identical internal states of photon pairs (i.e., the same for different  input ports $k$ with    $|\phi_j\rangle = |\psi_j\rangle$),     such internal states  still  lead to partial distinguishability, similar to  mixed states of single photons  \cite{SuffCond,DTh}.

With the definition in Eq. (\ref{E7}), in  the degenerate case, by  repeating the steps performed in   Eq. (\ref{E4}) of the previous section   we obtain  
\begin{eqnarray}
\label{E8}
 \!\!\! &&  \prod_{k=1}^N|r_k) =  \mathcal{Z}\sum_{n=0}^\infty \binom{2n}{n}^\frac12 
  \hat{S}_{2n} \!\left[\sum_{k=1}^N\frac{r_k}{2} |k\rangle|k\rangle\otimes |\Phib^{(2)}_k\rangle \right]^{\!\otimes n}\nonumber\\
  \!\!\! &&  \mathcal{Z} \equiv \prod_{k=1}^N Z_k  = \prod_{k=1}^N\prod_{j=1}^\infty \left(1- r_k^2p^{(k)}_j\right)^{\frac14}.
\end{eqnarray}
In Eq. (\ref{E8}),   according to  our splitting of a photon pair state into the  tensor product  of the operating and internal modes (explicitly indicated also by  ``$\otimes$"), the action of the particle  permutation  operator  $\hat{P}_\sigma$ in  the projector  $\hat{S}_{2n}$ (Eq. (\ref{E3}))  is split accordingly  $\hat{P}_\sigma\to \hat{P}_\sigma\otimes \hat{P}_\sigma$, where  the  factors  act  on the operating and the internal modes, respectively.

Now, observe that the $2n$-particle  state to the right of the  symmetrization operator $\hat{S}_{2n}$  in Eq. (\ref{E8}) already has some    symmetry by construction. Indeed,  if we permute the two photons in a photon pair from the same    source (i.e., with the same index $k$), which amounts to permuting    coinciding internal states,  or the photon  pairs,  i.e., the states  $|k\rangle|k\rangle\otimes |\Phib^{(2)}_k\rangle$, the  mentioned $2n$-particle  state  does not change.  Hence,  instead of applying the whole symmetric group $S_{2n}$ of $2n!$ permutations to symmetrize such a  state, one  can use instead only the set of  permutations which are \textit{different} matchings of $2n$ objects. Let us denote  the    $(2n-1)!!$  different   matchings by $\mathcal{M}_{2n}$.   The elements of $\Mc_{2n}$  can be enumerated by  vector-index  $ \alphb\equiv (\alpha_1,\ldots, \alpha_{2n})$,  
\be
\alphb\in \mathcal{M}_{2n}: \quad \alpha_{2i-1}< \alpha_{2i+1},  \quad \alpha_{2i-1} < \alpha_{2i}, 
\en{Vind}
where   the $i$th  matching   pair  is  $(\alpha_{2i-1},\alpha_{2i})$ (observe that $\alpha_1=1$).    For example, $\Mc_4$ consists of just three  permutations
\be
\Mc_4 = \{I, (2,3), (2,3,4) \},
\en{M_4}
where  $I$ stands for the trivial permutation and  $(i_1,i_2,\ldots, i_k)$ for the  cyclic permutation  $i_1\to i_2\to\ldots \to i_k\to i_1$ (thus, for instance,  $(2,3)$ is the transposition of two elements).  We have, for example, $\alphb_{(2,3)} = (1,3,2,4)$ with the two pairs being $(1,3)$ and $(2,4)$.     Below  the greek letters ``$\alpha$" and  ``$\beta$"  are used exclusively to enumerate matchings of $2n$ elements.

Denote  also by vector $\alphb$ the   permutation of $2n$ elements  $\quad \alphb (k) \equiv   \alpha_k$ corresponding to   the matching.    We can   project an arbitrary permutation   $\sigma\in S_{2n}$    on $\Mc_{2n}$.     For such a projection we will use the notation $\Mc(\sigma) = \alphb$.  Indeed, let us   expand  $\sigma$ as follows
\be
\sigma =   \pi \left( t_1\otimes \ldots \otimes t_n \right)   \alphb,
\en{Perm} 
where  $\pi \in S_n$ permutes  $n$ pairs, and $t_i \in S_2$ permutes the two elements of  the $i$th pair.  By Eq. (\ref{Perm}) the symmetrization projector  $\hat{S}_{2n}$  in  Eq. (\ref{E3}) can be factored as follows 
\begin{eqnarray}
\label{Proj}
&& \hat{S}_{2n} =    \hat{S}^{(pair)}_n\hat{S}_2^{\otimes n}\hat{\Mc}_{2n} ,  \nonumber\\
&&  \hat{\Mc}_{2n} \equiv \frac{1}{(2n-1)!!}\sum_{\alphb\in \Mc_{2n}} \hat{P}_\alphb, 
 \end{eqnarray}
where   the identity  $(2n)! = 2^n n! (2n-1)!!$ was used.  Since the  set of matchings $\Mc_{2n}$ is not a group (see  appendix \ref{appB}), the operator  $\hat{\Mc}_{2n}$ is not a projector (the other  operators in Eq. (\ref{Proj}) are projectors). 

 Now the crucial step is that the first  two factors   in the expression for $\hat{S}_{2n}$ in Eq. (\ref{Proj})  can   be dropped in Eq. (\ref{E8}),  since they have no effect (observe that in Eq. (\ref{E3}) the inverse permutation is applied to  indices of states   in a tensor product, thus a composition of permutations is applied in the reverse order).   

Similar as the above and in  analogy to  Eq. (\ref{E6}) of the previous section, in the non-degenerate case  we obtain 
\begin{eqnarray}
\label{E9}
 \!\!\! \prod_{k=1}^N|\tilde{r}_k) &\!=\!& \widetilde{ \mathcal{Z}}\sum_{n=0}^\infty \binom{2n}{n}^\frac12  
  \hat{S}_{2n} \!\left[\sum_{k=1}^Nr_k |k\rangle|N\!+\!k\rangle\otimes| \widetilde{\Phib}^{(2)}_k\rangle \right]^{\!\otimes n}\nonumber\\
\widetilde{ \mathcal{Z}} &\!=\!& \prod_{k=1}^N \widetilde{Z}_k= \prod_{k=1}^N\prod_{j=1}^\infty \left(1- r_k^2p^{(k)}_j\right)^{\frac12}.
\end{eqnarray}
In this case the quantum state to which the symmetrization projector $\hat{S}_{2n}$ is applied is not symmetric with respect to permutations within each photon pair due to different  inputs  modes ($|k\rangle$ and $|N+k\rangle$).  Therefore only the first  factor $ \hat{S}^{(pair)}_n$  in Eq. (\ref{Proj})    has no effect   (the state to which it is applied is the  $n$th power of  a two-photon state). The  last  two factors  in Eq.  (\ref{Proj})  select $n$   ``matchings with order"  of $2n$ objects,  where the order of the two objects   in each pair matters.   Therefore, to     reduce the projector   $\hat{S}_{2n}$ to a matching operator  in the non-degenerate case we need to introduce   the set $\widetilde{\Mc}_{2n}$  of $2^n(2n-1)!!$    matchings with order. These are   given by  the permutations $ \widetilde{\alphb} = \left( t_1\otimes \ldots \otimes t_n \right)  \alphb$, see  Eq. (\ref{Perm}), i.e.,    
 \be
\widetilde{\alphb}\in \widetilde{\Mc}_{2n}: \quad \mathrm{min}(\tilde{\alpha}_{2i-1},\tilde{\alpha}_{2i}) <  \mathrm{min}(\tilde{\alpha}_{2i+1},\tilde{\alpha}_{2i+2}).
\en{Vind2}

When   the internal modes of the photons with  two orthogonal polarizations coincide, the two-photon internal state  becomes the same   as  in  the  degenerate case,   $|\widetilde{\Phib}^{(2)}_{k}\rangle = | {\Phib}^{(2)}_{k}\rangle$.  If we apply  the projector  $\hat{S}_2^{\otimes n}$  to such an  input state, then the  two-photon term for each pair of inputs  in Eq. (\ref{E9}) is replaced  by  its projector on the symmetric two-photon state, 
\begin{eqnarray}
\label{AD1}
&&  \hat{S}_2 |k\rangle|N\!+\!k\rangle\otimes| {\Phib}^{(2)}_k\rangle \nonumber\\
&&=   \left[\frac{|k\rangle|N\!+\!k\rangle + |N+k\rangle| k\rangle}{2}\right]\otimes
 | {\Phib}^{(2)}_k\rangle.
\end{eqnarray}
 In this way such   $N$ special  non-degenerate squeezed states  at input ports $k=1,\ldots,2N$  of an  interferometer $U$ becomes equivalent to     $N$ pairs of degenerate squeezed states, where the  factor $1/2$   goes to the   squeezing parameter  $r_k\to r_k/2$ (compare Eqs. (\ref{E8}) and  (\ref{E9}))   with   the $k$th pair being   launched to the   inputs $k$ and $N+k$.  The  equivalence is realized  by   $N$  polarization-to-propagation mode beamsplitters  of   Eq. (\ref{PBS}), with the  output ports of   beamsplitter $k$ connected to inputs $k$ and $N+k$ of the interferometer $U$ (see also subsection \ref{sec3A}).
  
Below we will give  derivations   for   the  output probability in the degenerate case and  only  give the respective   results in the non-degenerate case. 
The  internal states $|\Phib^{(2)}_k\rangle$  Eq.~(\ref{E7})  are not affected by the  interferometer  and are not resolved at the detection stage (by our definition of the internal modes).  Introduce  the  sequence of  output ports $1\le l_1\le \ldots \le l_{2n}\le M$, one for each detected photon in an output configuration  $\mb = (m_1,\ldots, m_M)$, see Fig. \ref{F1}.  The photon counting   detection without internal state resolution  is given by  the following POVM operator (see  Refs. \cite{SuffCond,DTh})
\be
\hat{\Pi}_\mb = \frac{1}{\mb!} \sum_{j_1=1}^\infty \ldots \sum_{j_{2n}=1}^\infty\left[\prod_{i=1}^{2n} \hat{b}^\dag_{l_i \phi_{j_i}}\right]|0)(0|\prod_{i=1}^{2n} \hat{b}_{l_i \phi_{j_i}} ,
\en{E11}
where $\mb! = m_1!\ldots m_M!$ and the summation is over some (arbitrary) basis  $|\phi_j\rangle$ of the internal modes.  Using    Eq. (\ref{E3}),  the POVM operator  can be also cast in the first-order quantization representation.   We obtain   
\begin{eqnarray}
\label{E15}
  \hat{\Pi}_\mb = \frac{(2n)!}{\mb!} \hat{S}_{2n}   \left[{\prod_{i=1}^{2n}}  |l^{(out)}_i\rangle\langle l^{(out)}_i| \otimes  \openone  \right] \hat{S}_{2n}  ,
     \end{eqnarray} 
where    $\openone $ is the identity operator  in the subspace of   internal modes.   We can    omit the two projectors  $\hat{S}_{2n}$ from the expression in Eq. (\ref{E15}), since the quantum state  to which $ \hat{\Pi}_\mb$  will be applied   is already symmetric.

To simplify the  presentation below, for   each  output configuration  $\mb=(m_1,\ldots,m_M)$, such that  $|\mb|=2n$,  let us introduce the $M\times 2n$ matrix $\U$,  derived from the interferometer matrix $U$ Eq. (\ref{E10})  by taking  the rows $k=1,\ldots, M$  and  the columns $ l_1,   \ldots,  l_{2n}$  (generally, a multi-set),  i.e., we set 
\be
\U_{ki} \equiv U_{k,l_i}.
\en{MU}  
The   probability $p_\mb$ of detecting $2n$ photons in an output configuration $\mb$ is given by  the average of the detection operator (\ref{E15}) on   the quantum state (\ref{E8}).    Using  that     the matching  operator $\hat{\Mc}_{2n}$  Eq. (\ref{Proj}) can replace the symmetrization projector $\hat{S}_{2n}$ in Eq. (\ref{E8}),   we arrive at the following result 
(see details in appendix \ref{appA}) 
\begin{eqnarray} 
 \label{E17d}
&& p_\mb =   \mathrm{Tr}\left\{ \hat{\Pi}_\mb   \prod_{k=1}^N|r_k)  (r_k|  \right\} \nonumber\\
&& =  \frac{p_\mathbf{0}}{\mb!} \sum_{\alphb\in\Mc_{2n}} \sum_{\betb\in\Mc_{2n}}  \sum_{k_1=1}^N \ldots \sum_{k_n=1}^N  \sum_{k^\prime_1=1}^N \ldots \sum_{k^\prime_n=1}^N \nonumber\\
&& \times \left[ \prod_{i=1}^n  r_{k_i}  \U^*_{k_i,{\alpha_{2i-1}}} \U^*_{k_i, {\alpha_{2i}}}  r_{k^\prime_i}  \U_{ k^\prime_i, {\beta_{2i-1}}} 
\U_{k^\prime_i, {\beta_{2i}}} \right]\nonumber\\
    && \times \langle \Phib^{(2)}_{k_1}|  \ldots  \langle \Phib^{(2)}_{k_n}| \hat{P}^\dag_{\alphb}\hat{P}^{}_{\betb}|\Phib^{(2)}_{k^\prime_1}\rangle \ldots |\Phib^{(2)}_{k^\prime_n}\rangle,
\end{eqnarray}
where the  factor  $ p_\mathbf{0} \equiv  \Zc^2$ is  the probability to detect zero photons.
 The output probability  in  Eq.  (\ref{E17d}) has a  similar form  to  the output probability in   quantum  interference of partially distinguishable  single photons    \cite{SuffCond,DTh}, where   the  double sum  over all possible permutations in the product of the quantum amplitude   and  the  complex conjugate amplitude is  weighted  by a function of the  relative permutation.

  For $N$ non-degenerate squeezed states at  interferometer  input, the  output probability, $\tilde{p}_\mb$, can   be easily  recovered  from Eq. (\ref{E17d}) by replacing the normalization $\Zc^{(d)} \to  \widetilde{\Zc}$, the  internal states $|\Phib^{(2)}_k\rangle \to |\widetilde{\Phib}^{(2)}_k\rangle$,   the second instance of matrix element $U_{kl}$ with  $U_{N+k,l}$, and   the summation over matchings $\Mc_{2n}$ Eq. (\ref{Vind})   by  that over matchings with order $\widetilde{\Mc}_{2n}$  Eq. (\ref{Vind2}). We have  $ \widetilde{p}_\mathbf{0} =  \widetilde{\Zc}^2$ and 
\begin{eqnarray} 
 \label{E17nd}
&& \!\! \tilde{p}_\mb =  \frac{ \widetilde{p}_\mathbf{0} }{\mb!} \sum_{\alphb\in\widetilde{\Mc}_{2n}} \sum_{\betb\in\widetilde{\Mc}_{2n}}  \sum_{k_1=1}^N \ldots \sum_{k_n=1}^N  \sum_{k^\prime_1=1}^N \ldots \sum_{k^\prime_n=1}^N \nonumber\\
&& \times \left[ \prod_{i=1}^n  r_{k_i}  \U^*_{k_i,{\alpha_{2i-1}}} \U^*_{N+k_i ,{\alpha_{2i}}}  r_{k^\prime_i}  \U_{ k^\prime_i, {\beta_{2i-1}}} 
\U_{N+k^\prime_i, {\beta_{2i}}} \right]\nonumber\\
    && \times \langle \widetilde{\Phib}^{(2)}_{k_1}|  \ldots  \langle \widetilde{\Phib}^{(2)}_{k_n}| \hat{P}^\dag_{\alphb}\hat{P}^{}_{\betb}|\widetilde{\Phib}^{(2)}_{k^\prime_1}\rangle \ldots |\widetilde{\Phib}^{(2)}_{k^\prime_n}\rangle.
\end{eqnarray}

  Below, we will analyze the probabilities  in Eqs. (\ref{E17d}) and (\ref{E17nd})  in some special cases. 

%%%%%%%%%%%%%%%%%%%%%%%%%%%%%%
\subsection{The ideal case:  output  amplitude as   Hafnian}
\label{sec3A}

The ideal case of   the  interference with  squeezed states  can be    defined by the absence of any dependence on the   internal modes, similar as  with single photons \cite{SuffCond,DTh}. 
There is  no dependence on the internal modes  in  the output  probability in Eq. (\ref{E17d})  when the  matching operator does not affect the tensor product of the internal states of photon pairs.  This occurs when the internal states   coincide    and, moreover,  there is just  one internal mode, i.e., $|\Phib^{(2)}_k\rangle = |\phi_1\rangle|\phi_1\rangle$.  In this case, from Eq. (\ref{E17d}) we get  the  well-known expression  \cite{hafPhys,GBS2} for the output  probability from  the Gaussian states $  \mathring{p}_\mb$:
    \begin{eqnarray}
 \label{E18}
 &&  \mathring{p}_\mb = \frac{   \mathring{p}_\mathbf{0} }{\mb!} \left|  \sum_{\alphb\in\Mc_{2n}}\prod_{i=1}^n  \sum_{k_i=1}^N  \U_{k_i,{\alpha_{2i-1}}} r_{k_i} \U_{k_i, {\alpha_{2i}}}\right|^2\nonumber\\
 && =   \frac{  \mathring{p}_\mathbf{0} }{\mb!} \biggl| \sum_{\alphb\in \Mc_{2n}} \prod_{i=1}^n\mathcal{A}_{{\alpha_{2i-1}},{\alpha_{2i}}}\biggr|^2,
\end{eqnarray}
where $  \mathring{p}_\mathbf{0}   =   \prod\limits_{k=1}^N\  \left(1- r_k^2 \right)^{\frac14}$  and the $2n$-dimensional symmetric matrix $\mathcal{A}$ is defined as follows
\be
\mathcal{A}_{ij} = \sum\limits_{k=1}^N\U_{ki}r_k\U_{kj} =  \sum\limits_{k=1}^N U_{kl_i}r_kU_{kl_j} .
\en{Ad}
The sum over matchings  $\sum_{\alphb\in \Mc_{2n}} \prod_{i=1}^n\mathcal{A}_{{\alpha_{2i-1}},{\alpha_{2i}}}$  in the last row in Eq. (\ref{E18})   is  called   Hafnian of  a symmetric matrix $\mathcal{A}$ \cite{hafPhys}. 
 
In the non-degenerate case  the  conditions for the ideal interference   require  the same   internal mode $|\phi\rangle =|\psi\rangle$ for the two polarizations.  From Eq. (\ref{E17nd}) one can get the following expression for the probability   
\begin{eqnarray}
 \label{E20}
   \mathring{\widetilde{p}}_\mb    =   \frac{ \mathring{\widetilde{p}}_\mathbf{0}}{\mb!} \biggl| \sum_{\alphb\in \widetilde{\mathcal{M}}_{2n}}  
  \prod_{i=1}^n\mathcal{\tilde{A}}_{{\alpha_{2i-1}},{\alpha_{2i}}}\biggr|^2 \nonumber\\
   \end{eqnarray}
where  $\mathring{\widetilde{p}}_\mathbf{0}  =\mathring{\widetilde{\Zc}} =  \prod\limits_{k=1}^N\  \left(1- r_k^2 \right)^{\frac12}$ and 
\be
\mathcal{\tilde{A}}_{ij} = \sum\limits_{k=1}^N\U_{ki}r_k\U_{N+k,j} =  \sum\limits_{k=1}^N U_{kl_i}r_kU_{N+k,l_j}.
\en{And}
Only the symmetric part of the matrix in Eq. (\ref{And}) contributes  in Eq. (\ref{E20}) due to the  sum    $\mathcal{\tilde{A}}_{\alpha_{2i-1},\alpha_{2i}}+\mathcal{\tilde{A}}_{\alpha_{2i},\alpha_{2i-1}}$.  Hence,  the summation  over the ordered matchings $\widetilde{\mathcal{M}}_{2n}$ Eq. (\ref{Vind2}) can be reduced  to that over the usual matchings  ${\mathcal{M}}_{2n}$ Eq. (\ref{Vind}), while  retaining only  the symmetric part of $\mathcal{\tilde{A}}$.  The probability  in Eq. (\ref{E20}) takes the form  
\be
 \mathring{\widetilde{p}}_\mb =  \frac{ \mathring{\widetilde{p}}_\mathbf{0}}{\mb!} \biggl| \sum_{\alphb\in  {\mathcal{M}}_{2n}}  
  \prod_{i=1}^n\mathcal{\tilde{A}}^{(s)}_{{\alpha_{2i-1}},{\alpha_{2i}}}\biggr|^2, 
\en{E20A}
where $\mathcal{\tilde{A}}^{(s)}_{ij} = \mathcal{\tilde{A}}_{ij} +\mathcal{\tilde{A}}_{ji} $.  

The expression in Eq. (\ref{E20A}) is equivalent  to that of Eq. (\ref{E18}) for a different interferometer $U^\prime$.   Indeed, consider the  matrix 
$\mathcal{\tilde{A}}$  with the following   matrix elements in Eq. (\ref{And}):
\be
U_{kl} \equiv \frac{U^\prime_{kl} +i U^\prime_{N+k,l} }{\sqrt{2}},\quad  U_{N+k,l} \equiv \frac{ U^\prime_{kl} -i U^\prime_{N+k,l} }{\sqrt{2}}.
\en{Uprime}
Then   $\mathcal{\tilde{A}}^{(s)}$  becomes
\be
\mathcal{\tilde{A}}^{(s)}_{ij} =  \mathcal{\tilde{A}}_{ij} +\mathcal{\tilde{A}}_{ji} = \sum_{k=1}^{2N} r_kU^\prime_{k,l_i}U^\prime_{k,l_j}. 
\en{Asym}
The transformation in Eq. (\ref{Uprime}) is the  interferometer  $U^\prime$ preceded  by  $N$ auxiliary polarization-to-propagation mode beamsplitters   as in  Eq. (\ref{PBS})  of section \ref{addsec2} (recall that in the non-degenerate case    the input port index includes also the polarization mode),  where  beamsplitter $k$  receives as the  inputs the two polarization modes of the $k$th non-degenerate squeezed state and  is connected  to inputs $k$ and $N+k$ of the interferometer $U^\prime$.     The auxiliary beamsplitters transform  $N$ non-degenerate squeezed states   into $2N$ degenerate ones in the same polarization mode,   as  in Eq. (\ref{Effect}).  One can interpret  $  \mathring{\widetilde{p}}_\mathbf{0} $ of Eq. (\ref{E20A})  as the probability to detect zero photons for  the above $2N$ degenerate squeezed states with the squeezing parameters $r_k$ and $r_{N+k}\equiv r_k$, $k=1,\ldots, N$.  Therefore, the probability in Eq. (\ref{E20A})  has the form of that in Eq. (\ref{E18}) where  $\mathcal{A}_{ij}\equiv \mathcal{\tilde{A}}^{(s)}_{ij}$ of Eq. (\ref{Asym}), and corresponding to $N$ pairs of  squeezed states, where  pair $k$ with the squeezing parameter $r_k$ is  launched to inputs $k$ and $N+k$ of the new interferometer $U^\prime$.

%%%%%%%%%%%%%%%%%%%%%%%%%%%%%%
\subsection{Identical multi-mode  internal states }
\label{sec3B}

Consider now the case of coinciding multi-mode internal states of photon pairs in  different input ports of the interferometer $U$,
 \be
 |\Phib^{(2)}_k\rangle =  \sum_{j=1}^\infty \sqrt{p}_j|\phi_j \rangle|\phi_j\rangle \equiv |\Phib^{(2)}\rangle, \quad k = 1,\ldots, N.  
\en{E22}
This model  allows for further analysis, on the one hand, and, on the other hand, applies to the recent experiment  on Gaussian boson sampling \cite{ExpGBS2}, where  a coherent splitting of a  single  pump source was used to generate  the squeezed states.  

The expression in Eq. (\ref{E17d})  can now be further simplified as follows. First,  we can  perform the summations  over $k_i$ and $k^\prime_i$ with the result
 \begin{eqnarray}
 \label{E17dA}
   p_\mb &=&     \frac{ p_\mathbf{0}}{\mb!}  \sum_{\alphb\in \Mc_{2n}} \sum_{\betb\in \Mc_{2n}}
 \prod_{i=1}^n\mathcal{A}^*_{{\alpha_{2i-1}},{\alpha_{2i}}} \mathcal{A}_{{\beta_{2i-1}},{\beta_{2i}}} \nonumber\\
 & & \times   \langle \Phib^{(2)}|^{\otimes n}   \hat{P}_{\alphb^{-1}\betb} |\Phib^{(2)}\rangle^{\otimes n},
\end{eqnarray}
where $\mathcal{A}$ is given by Eq. (\ref{Ad}).  Second,    the matching operator in Eq. (\ref{E17dA})  now acts on the internal state   $|\Phib^{(2)}\rangle^{\otimes n}$  invariant with  respect to permutations of the two-photon states $|\Phib^{(2)}\rangle$  and with respect to transposition  of   two identical internal modes $|\phi_j\rangle|\phi_j\rangle$   of  each  photon pair  (i.e., the  same  symmetry  as in the input state of Eq. (\ref{E8})). Therefore, we can replace the relative  permutation $\alphb^{-1}\betb\in S_{2n}$ in Eq. (\ref{E17dA})    by  its projection on the matchings $\Mc(\alphb^{-1}\betb)\in \Mc_{2n}$.   As the result,  we have to consider   only  the average of a  matching operator on the internal state of $2n$ photons, i.e.,  study a function on $\Mc_{2n}$ defined as 
 \be
J(\alphb) \equiv  \langle \Phib^{(2)}|^{\otimes n} \hat{P}_{\alphb}|\Phib^{(2)}\rangle^{\otimes n}.
 \en{E23}

%%%%%%%%%%%%
\subsubsection{Disjoint cycle decomposition of a matching}
\label{secCycle}

\begin{figure}[h]
\begin{center}
      \includegraphics[width=.35\textwidth]{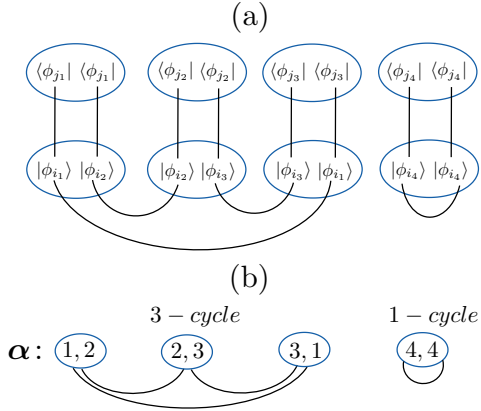} 
     \caption{\textbf{The matching cycles.} \textbf{(a)}  The ovals represent  states of  photon pairs. The action of    $\hat{P}_{\alphb}$ on the ket vectors in   $|\Phib^{(2)}\rangle^{\otimes 4}$   is shown in the bottom ovals (by the curvy lines), whereas the top ovals give the corresponding bra vectors in the inner product $\langle \Phib^{(2)}|^{\otimes 4}\hat{P}_{\alphb}|\Phib^{(2)}\rangle^{\otimes 4}$.  The curvy lines connect  the same  ket vectors, whereas the vertical  lines  connect   the   ket vectors to the respective bra vectors in the inner product.        \textbf{(b)}~ The  cycle decomposition of the corresponding  matching   $\alphb\in\Mc_{8}$  (in the  ovals)    {acting  on the  double-set} $(1,1,2,2,3,3,4,4)$, composed of the indices of    internal modes.    There is  a $3$-cycle and a   $1$-cycle (fixed point).   \label{F2} }
   \end{center}
\end{figure}

At this point is it necessary to  introduce   what will be called the ``cycle decomposition"  of   a matching  $\alphb\in \Mc_{2n}$ acting  on a double-set $\mathbf{x} \equiv (1,1,\ldots,n,n)$. The double-set  appears here  due to coinciding  indices of the internal states ($|\phi_j\rangle$) in each photon pair. Let us rearrange    a matching  of the double-set,  $ \alphb(\mathbf{x}) \equiv  (x_{\alpha_1}, x_{\alpha_2}, \ldots ,x_{\alpha_{2n}})$, as follows. Starting from the first pair $(j_1,j_2)$, with $j_1=1$ (since $x_{\alpha_1}=1$) and $j_2 = x_{\alpha_2}$,   we look for the next pair containing $j_2$, say $(j_2,j_3)$  (permuting  the two elements in  such a pair, if necessary, to put  $j_2$ on the first place).   We continue by looking for the pair containing now $j_3$, etc, until we have come to a pair with $j_{k+1}=j_1=1$, for some $1\le k\le n$, i.e., we  end up with a  cycle $\nu$ of length $k$, or $k$-cycle:
\be
\nu \equiv\left\{ (j_1,j_2), (j_2,j_3),\ldots, (j_{k},j_1)\right\}.
\en{Cycle}
(Observe that  a matching  cycle of length $k$  contains $k$   pairs of elements and that each element is repeated.)
Then, starting from the smallest $j\notin \nu$ of Eq. (\ref{Cycle}), quite similarly we end up with another cycle, $\nu^\prime$,  starting and ending with   $j$. We continue until all the elements of $\mathbf{x}$ are arranged in such disjoint cycles (i.e., not having elements in common).  In this way, a  matching permutation, acting on a double-set,   is cast as  a product of    disjoint cycles.   

Denote by  $C_k(\alphb)$  the total number of  $k$-cycles   Eq. (\ref{Cycle}) in a matching $\alphb\in \Mc_{2n}$. The  numbers $C_1,\ldots, C_n$ satisfy the obvious  constraint $\sum_{k=1}^n kC_k =  n$. 

A  permutation  $\alphb$ and the  inverse permutation $\alphb^{-1}$  correspond to the same cycle structure $(C_1,\ldots, C_n)$. Since  permutation $\alphb$ is cast as   the  product of disjoint cycles, consider just a single cycle $\nu$,   Eq. (\ref{Cycle}).   The  cycle $\nu$  of Eq. (\ref{Cycle}) can be   obtained by application     of   the following cyclic permutation in $S_n$  of length $k$
\be
j_{k} \to j_{k-1} \to \ldots \to j_{1} \to j_{k} 
\en{permcycle} 
to    the second element   in each pair in  the trivial matching   $\{ (j_1,j_1), (j_2,j_2),\ldots, (j_{k},j_k)\}$. 
It is easy to see that the inverse  permutation $j_{1} \to j_{2} \to \ldots \to j_{k} \to j_{1}$   results in the same cycle $\nu$ (with  the pairs  permuted; see also   appendix \ref{appC}).  Hence, $C_k(\alphb^{-1}) = C_k(\alphb)$ for all $k=1,\ldots, n$. 

%%%%%%%%%%%%
\subsubsection{The output probability      }
\label{secProb}

Consider now  the matching  operator $\hat{P}_\alphb$ in Eq. (\ref{E23}).  It  can be  factorized  into a product of operators of  disjoint cycles $\hat{P}_\alphb = \prod_i \hat{P}_{\nu_i}$.  The inner product in Eq. (\ref{E23}) factorizes accordingly. Due to orthogonality of the internal modes, $\langle\phi_j|\phi_{j^\prime}\rangle  = \delta_{jj^\prime}$, the operator  $\hat{P}_{\nu_i}$ of a single cycle  has a non-zero contribution  to the average in Eq. (\ref{E23})   only when in   each  inner product   of the cycle $\nu_i$  the corresponding bra and ket $\phi$-states  coincide (i.e., have the same index). As seen from Fig. \ref{F2}, this necessitates that    all  the   bra and ket  $\phi$-states  within each independent    cycle  $\nu_i$ coincide  (have the same index). Therefore,  a $k$-cycle contributes the  factor  $\sum_j p^k_j$ to $J(\alphb)$, since there are $2k$ products of bra and ket $\phi$-states, each weighed by $\sqrt{p_j}$ see Eq. (\ref{E22}).  In other words,  only  the \textit{diagonal part}   of the photon pair   state  $ |\Phib^{(2)}\rangle \langle  \Phib^{(2)}|$ contributes to output probabilities, where every $k$-cycle contributes  a  factor equal to the trace of the $k$th power of the diagonal part of the photon pair state. 
  
 From the above discussion  the distinguishability function $J(\alphb)$ Eq. (\ref{E23}) becomes
\be
J(\alphb) =    \prod_{k=2}^n \Biggl( \sum_{j=1}^\infty p^k_j\Biggr)^{C_k(\alphb)},
\en{E25}
where  the omitted factor due to   $1$-cycles (fixed points) is equal to $1$. The expression in  Eq. (\ref{E25})   reminds a similar expression for the distinguishability function of   single  photons,  in the case when each   photon is in   the same mixed internal state \cite{SuffCond,DTh,CollPh}. In the latter case  the  cyclic  permutations of photons  also contribute as independent factors to the distinguishability function.  However, there are two new elements here: the cycles rearrange the  photon pairs, not single photons, and, therefore, the permutation group  $S_n$ is replaced by the set of matchings $\Mc_{2n}$.   
  
 With the  distinguishability function of  Eq. (\ref{E25}) the output probability becomes
  \begin{eqnarray}
 \label{E17dB}
  p_\mb &=&     \frac{p_\mathbf{0}}{\mb!}  \sum_{\alphb\in \Mc_{2n}} \sum_{\betb\in \Mc_{2n}}
   \prod_{k=2}^n \Biggl( \sum_{j=1}^\infty p^k_j\Biggr)^{C_k(\alphb^{-1}\betb)} \nonumber\\
 & & \times  \prod_{i=1}^n\mathcal{A}^*_{{\alpha_{2i-1}},{\alpha_{2i}}} \mathcal{A}_{{\beta_{2i-1}},{\beta_{2i}}},
  \end{eqnarray}
where $C_k(\alphb^{-1}\betb)$ is the number of $k$-cycles  in the disjoint  cycle decomposition of the matching  $\Mc(\alphb^{-1}\betb)\in \Mc_{2n}$.  Since $C_k(\alphb^{-1}\betb) = C_k(\betb^{-1}\alphb)$,  only the real part of  the product of  matrix elements contributes to the probability    in Eq. (\ref{E17dB}). 

A similar expression for the output probability can be derived in the non-degenerate case,  when   the internal modes for  the two polarizations are the same. In this case  $|\widetilde{\Phib}^{(2)}_k\rangle = |\Phib^{(2)}\rangle$ with $|\Phib^{(2)}\rangle$ of Eq. (\ref{E22}) (recall that the polarizations are excluded from the internal states).  Then,  by similar arguments as in section \ref{sec3A} one can show that only the symmetric part of the matrix given by  Eq. (\ref{And}) contributes to the probability and the result is equivalent to that of the degenerate case with the matrix of Eq. (\ref{Asym}).  Taking this into account,  we obtain
\begin{eqnarray}
 \label{E17ndB}
 \tilde{p}_\mb &=&     \frac{ \widetilde{p}_\mathbf{0}}{\mb!}  \sum_{\alphb\in  {\Mc}_{2n}} \sum_{\betb\in  {\Mc}_{2n}}
   \prod_{k=2}^n\Biggl( \sum_{j=1}^\infty p^k_j\Biggr)^{C_k(\alphb^{-1}\betb)}\nonumber\\
 & & \times\prod_{i=1}^n\mathcal{A}^*_{\alpha_{2i-1},\alpha_{2i}} \mathcal{ A}_{{\beta_{2i-1}} ,{\beta_{2i}}}  
 \end{eqnarray}
where $\mathcal{A}_{ij}\equiv \mathcal{\tilde{A}}^{(s)}_{ij}$  of   Eqs. (\ref{Uprime})-(\ref{Asym}). 

%%%%%%%%%%%%%%
\subsubsection{Example: Probability to detect  four photons}
Consider the probability to  detect just    four photons at  interferometer output, i.e. $n=2$.   For $n=2$ there are only 1-cycles (i.e., fixed points) and $2$-cycles, thus   Eq. (\ref{E17dB}) depends only on the number of 2-cycles $C_2(\alphb^{-1}\betb)$.  
The three permutations in the set  $\Mc_{4}$  are given in Eq. (\ref{M_4}). Let us denote   $\mu_2 = (2,3)$ and $\mu_3 = (2,3,4)$ (respectively, the transposition of $2$ and $3$ and the cycle $2\to3\to4\to2$). Observing that $(2,3)^{-1} = (2,3)$ and $(2,3)(2,3,4) = (2,4)$, we have their action on $\{1,2,3,4\}$:
\begin{eqnarray}
\label{M_4rule}
&& \mu_2\left(\begin{array}{c}1\\ 2\\ 3\\ 4\end{array}\right)=\left(\begin{array}{c}1\\ 3\\ 2\\ 4\end{array}\right), \quad \mu_3 \left(\begin{array}{c}1\\ 2\\ 3\\ 4\end{array}\right) = \left(\begin{array}{c}1\\ 4\\ 2\\ 3\end{array}\right) \nonumber\\
&&  \mu_2^{-1}\mu_3\left(\begin{array}{c}1\\ 2\\ 3\\ 4\end{array}\right)=\mu_3^{-1}\mu_2\left(\begin{array}{c}1\\ 2\\ 3\\ 4\end{array}\right)=\left(\begin{array}{c}1\\ 4\\ 3\\ 2\end{array}\right),\nonumber\\
&&  \mu_3^{-1}\left(\begin{array}{c}1\\ 2\\ 3\\ 4\end{array}\right) = \left(\begin{array}{c}1\\ 3\\ 4\\ 2\end{array}\right).
\end{eqnarray} 
 Now, let us  find the number of 2-cycles  of the relative permutations    acting on the double set $\{1,1,2,2\}$. In this case the   index $1\le i \le 4$   in Eq. (\ref{M_4rule}) points to  the $i$th  element    in the double set.      
One can represent the number of 2-cycles for the  nine    relative permutations  $\alphb^{-1}\betb$, $\alphb,\betb\in \{I,\mu_2,\mu_3\}$   by a matrix  
$\mathcal{C}_{\alphb,\betb}\equiv C_2(\alphb^{-1}\betb)$, where
\be
\mathcal{C} = \left(\begin{array}{ccc}  0 & 1 & 1\\1 & 0 & 1  \\ 1 & 1& 0 \end{array} \right) 
\en{M4table}
with the rows and columns corresponding  to $\alphb$ and $\betb$ in the order $(I, \mu_2, \mu_3)$.  
For instance, $C_2(\mu_3^{-1}I) =\mathcal{C}_{31} = 1$, which can be read from the action of $\mu_3^{-1}$   in Eq. (\ref{M_4rule}),  projected on the double set as above indicated,    and using the definition of a 2-cycle in Eq. (\ref{Cycle}). 

Using   Eq. (\ref{M4table}) into Eq.  (\ref{E17dB}) we  can now  write down the  probability to detect four  photons in an output configuration $\mb=(m_1,\ldots,m_M)$, $|\mb|=4$,  where at most four different  output  ports $1\le l_1\le l_2\le l_3\le l_4\le M$ are occupied by photons.  Recalling that for four photons  we have a  $4$-dimensional   matrix  $A_{l_i l_j}\equiv  \mathcal{A}_{ij} = \sum_{k=1}^N U_{kl_i}r_kU_{kl_j} $  and using the  definition of purity   $\mathbb{P}  = \sum_{j=1}^\infty p^2_j$ Eq. (\ref{SchmidtK}), we obtain  
\begin{eqnarray}
\label{prob_new}
 p_\mb &= &\frac{p_\mathbf{0} }{\mb!} \Biggl( |A_{l_1l_2}A_{l_3l_4}|^2 + |A_{l_1l_3}A_{l_2l_4}|^2  + |A_{l_1l_4}A_{l_2l_3}|^2   \nonumber\\
& +&  2\mathbb{P} \mathrm{Re}\Bigl\{ \left[A^*_{l_1l_3}A^*_{l_2l_4} +A^*_{l_1l_4}A^*_{l_2l_3}\right]A_{l_1l_2}A_{l_3l_4}  \nonumber \\
&+ & A^*_{l_1l_3}A^*_{l_2l_4}A_{l_1l_4}A_{l_2l_3}   \Bigr\}\Biggr).
\end{eqnarray}
In Eq.  (\ref{prob_new}) the first three terms    correspond to  $\mathcal{C}_{ii}$ for  $i=1,2,3$ and, in the real part,  the  next  two terms   to $\mathcal{C}_{1i}$ and $\mathcal{C}_{i1}$, with $i=2,3$,  while  the last term to $\mathcal{C}_{23}$ and $\mathcal{C}_{32}$.

Let us apply Eq. (\ref{prob_new}) to the interference  on a beamsplitter of   two degenerate squeezed   states with the squeezing parameters $r_1$ and $r_2$.  We have (without loss of generality)  
\be
U = \left(\begin{array}{cc} u & v\\ 
-v & u \end{array}\right), \quad u^2 +v^2 = 1,
\en{BS1}
where $0\le u,v\le 1$.   Therefore  
\be
A = \widetilde{U}\left(\begin{array}{cc} r_1& 0\\ 
0& r_2 \end{array}\right)U = \left(\begin{array}{cc} r_1u^2+r_2v^2 & (r_1-r_2)uv \\
(r_1-r_2) uv & r_1v^2+r_2u^2 \end{array}\right). 
\en{A}
For the five possible output configurations $\mb \in  \{(4,0), (0,4), (2,2), (3,1), (1,3)\}$   we have the following multisets of output port indices:
\begin{eqnarray*}
 &&(4,0) \rightarrow l_1=l_2=l_3=l_4=1, \\
 && (0,4) \rightarrow l_1=l_2=l_3=l_4=2, \\
 && (2,2) \rightarrow l_1=l_2= 1,\quad l_3=l_4=2,\\
 &&(3,1) \rightarrow l_1=l_2=l_3=1,\quad l_4=2,\\
 &&(1,3) \rightarrow l_1=1, \quad l_2=l_3= l_4=2.
\end{eqnarray*} 
Then  Eq. (\ref{prob_new}) gives:
\begin{eqnarray}
  p_{(4,0)} &= & \frac{p_{\mathbf{0}}}{4}A_{11}^4\left(\frac12 +\mathbb{P}\right),\quad  p_{(0,4)} =  \frac{p_{\mathbf{0}}}{4}A_{22}^4\left(\frac12 +\mathbb{P}\right),\nonumber\\
p_{(2,2)} &= & \frac{p_{\mathbf{0}}}{4}A^2_{11}A^2_{22}\left[1 + 2(1+\mathbb{P})   \frac{A_{12}^4}{A^2_{11}A^2_{22}} 
 +4 \mathbb{P} \frac{A_{12}^2}{A_{11}A_{22}}   \right]\nonumber\\
p_{(3,1)}& = & \frac{p_{\mathbf{0}}}{2}A^2_{11}A^2_{12}(1+ 2\mathbb{P}), \;\; p_{(1,3)}= \frac{p_{\mathbf{0}}}{2}A^2_{22}A^2_{12}(1+ 2\mathbb{P}).\nonumber\\
\label{BSprob}
\end{eqnarray}

Eq. (\ref{BSprob}) applies also to the interference of a single non-degenerate squeezed  state, where now   $r_1=r_2$, hence $A_{11} = A_{22}$, and $A_{12}=0$.  This is due to the equivalence   of the  probability formula  between  the   degenerate and non-degenerate cases, established in section \ref{sec3A}, when the internal modes are the same for the two polarizations in the non-degenerate case.    Eq. (\ref{BSprob})   also reproduces the four-photon interference probabilities on the balanced   beamsplitter $u=v=1/\sqrt{2}$     in the scheme of Ref.~\cite{ModeStrInfl}.    To show this,  let us perform post  selecting   on  the detection of  four photons at the output,  by  dividing the results in Eq. (\ref{BSprob}) by  the probability to detect exactly four photons  
\be
p(4) \equiv p_{(4,0)} + p_{(0,4)} + p_{(2,2)} = \frac{p_{\mathbf{0}}}{2}A_{11}^4\left(1 +\mathbb{P}\right),
\en{4PhProb}
 (in this case $p_{(1,3)} = p_{(3,1)}=0$).  For the   non-zero  conditional probabilities $P_{\mb}\equiv p_\mb/p(4)$  we get \cite{ModeStrInfl}
\be
P_{(4,0)}  = P_{(0,4)} =  \frac{1 +  2\mathbb{P}}{4+4 \mathbb{P}}, \quad P_{(2,2)} = \frac{1}{2+ 2\mathbb{P}}.
\en{Ref39}

The above four-photon interference on a beamsplitter  of two degenerate squeezed states, or of a single non-degenerate squeezed state  in  the scheme of  Ref.~\cite{ModeStrInfl},   can be used to  estimate the purity of the squeezed states. In  section \ref{sec4}  it will be shown that  the effect of distinguishability in the quantum interference with an arbitrary number of squeezed states, each having only  two common internal modes,  can also be expressed  only through    the purity.

%%%%%%%%%%%%%%%%%%%%%%%%%%%%%%
\subsection{Orthogonal  internal states}
\label{sec3C}

  We will use that the   output probability,  Eq.~(\ref{E17d}),  can be  also cast  in an equivalent  form   of a quantum average 
 \begin{eqnarray} 
 \label{E32}
&&    p_\mb =    \mathrm{Tr}\left\{ \hat{\Pi}_\mb   \prod_{k=1}^N|r_k)  (r_k|  \right\} = p_\mathbf{0}\binom{2n}{n}\frac{ (2n)!}{\mb!}  \nonumber\\ 
&&  \times \langle \Psib^{(2)}|^{\otimes n}   \hat{\Mc}^\dag_{2n}  \left[\prod_{i=1}^{2n} |l^{(out)}_i\rangle\langle l^{(out)}_i| \otimes  \openone  \right]    \hat{\Mc}_{2n}   |\Psib^{(2)}\rangle^{\otimes n},  \nonumber\\
\end{eqnarray}
where we have introduced an unnormalized $2$-particle state   
\begin{eqnarray}
\label{E33}
|\Psib^{(2)}\rangle \equiv     \sum_{k=1}^N  \frac{r_k}{2} |k \rangle |k\rangle \otimes |\Phib^{(2)}_{k}\rangle
 \end{eqnarray}
and observed that   $\left[2^n(2n-1)!!\right]^2=(2n)!\binom{2n}{n}$. 

Consider  now  the case of  mutually  orthogonal  internal states,  i.e.,  $\langle \Phib^{(2)}_{k^\prime}|\Phib^{(2)}_k\rangle = \delta_{k^\prime k}$, or, equivalently, $\langle \phi^{(k^\prime)}_{j^\prime}|\phi^{(k)}_j\rangle = 0$ for $k^\prime\ne k$, see Eq. (\ref{E7}).
Let us introduce  the   projectors $\hat{E}_1, \ldots, \hat{E}_N$  onto the internal Hilbert spaces  of  the  squeezed states:
\be
\hat{E}_k |\phi_j^{(k^\prime)}\rangle  = \delta_{k^\prime k}|\phi^{(k)}_j\rangle, \quad j=1,2,\ldots. 
\en{Ek}
 Then, without changing  the result, the following substitution can be made   in Eq. (\ref{E32})  
\begin{eqnarray}
\label{PiEk}
 |l^{(out)}_i\rangle\langle l^{(out)}_i| \otimes  \openone \to   |l^{(out)}_i\rangle\langle l^{(out)}_i| \otimes  \sum_{k=1}^N \hat{E}_k.
\end{eqnarray}
Moreover, since  each squeezed state contributes  pairs of photons,   only the terms   involving pairs of projectors   $\hat{E}_k$  contribute to the  output probability.     Thus we  can replace the  operator  in the square brackets in  Eq. (\ref{E32}) with the one accounting only for all possible occurrences of pairs  of  $\hat{E}_k$:
\begin{eqnarray}
\label{Detec}
&&  \!\!\! \prod_{i=1}^{2n}  \left[|l^{(out)}_i\rangle\langle l^{(out)}_i| \otimes  \sum_{k=1}^N \hat{E}_k \right] \to   \sum_{k_1=1}^N \ldots\sum_{k_n=1}^N \frac{1}{(2\nb)!}\\
&&  \times \sum_{\mu \in S_{2n}}  \left[ \prod_{i=1}^{2n} |l^{(out)}_i\rangle\langle l^{(out)}_i| \right] \otimes   \left[ \prod_{i=1}^n  \hat{E}_{k_{\mu(i)}}\otimes \hat{E}_{k_{\mu(n+i)}} \right], \nonumber
\end{eqnarray}
where the  pairs  $k_{i} = k_{n+i}$ from $\{1,2,\ldots, N\}$  with the occurrences $\nb = (n_1,\ldots,n_N)$ are   distributed over the output ports  by   permutation $\mu$ and $(2\nb)! = (2n_1)!\ldots(2n_N)!$  accounts for  multiple counting of the same  terms.  Observe that each term in Eq. (\ref{Detec})   describes the  detection of   photons and the information on the squeezed state each photon came from.  For instance, each such term is a collection of configurations  $\mb^{(k)} = (m^{(k)}_1,\ldots, m^{(k)}_M)$, $k=1, \ldots, N$,  such that $\mb^{(1)} + \ldots + \mb^{(N)} = \mb$,  where the output ports in the configuration $\mb^{(k)}$ correspond to   the tensor product with the same projector $\hat{E}_k$ in the internal subspace.   Since the normalization  factor is a product as well,  $p_\mathbf{0} = \Zc  = \prod_{k=1}^N Z_k = \prod_{k=1}^N {p_\mathbf{0}}_k$, see Eq. (\ref{E8}),   it is now obvious that, due to the mere   possibility of complete resolution of the  orthogonal  internal states of photons at the detection stage,  the output probability Eq. (\ref{E32})  is  a convex mixture of  products of the probabilities  from the  individual  squeezed states from different input ports:
\be
p_\mb =    \sum p^{(1)}_{\mb^{(1)}} \ldots  p^{(N)}_{\mb^{(N)}}, 
\en{EqProb}
  where the sum is constrained by $\mb^{(1)} + \ldots + \mb^{(N)} = \mb$.  In Eq. (\ref{EqProb})   we have denoted by $ p^{(k)}_{\mb^{(k)}}$  the probability to detect $2n_k = |\mb^{(k)}| $  photons from  the squeezed state at input port $k$ in the output configuration $\mb^{(k)}$, i.e.,  given by the same formula as in Eq. (\ref{E17dB})  with the substitutions $p_\mathbf{0}\to {p_\mathbf{0}}_k$ and  $ \mathcal{A}_{ij} \to   \mathcal{A}^{(k)}_{ij} \equiv \U_{ki}r_k\U_{kj}$.

%%%%%%%%%%%%%%%%%%%%%%%%%%%%%%
\subsection{General case of   internal  states  }
\label{sec3D}

Consider now the most general case of arbitrary  (different) internal  modes  of  photon pairs  at different input ports of    interferometer.    The  average of the relative matching operator $ \hat{P}^\dag_{\alphb}\hat{P}_{\betb}$ on the  product of the two-photon  internal  states in Eq. (\ref{E17d}) depends on  the sets  $k_1, \ldots, k_n$ and $k_1^\prime, \ldots, k^\prime_n$,  preventing summation  over these indices solely   in the product of the matrix elements of $U$.     The matching permutations $\alphb,\betb\in \Mc_{2n}$ Eq. (\ref{Vind}) do not form a group (see appendix \ref{appB}), and  the relative permutation  
$\alphb^{-1}\betb$ may differ from a matching  in  the   standard form of Eq. (\ref{Vind}), i.e.,  ${\alphb}^{-1} {\betb}  \notin \Mc_{2n}$.  Since the tensor product of the  internal  states to which such a permutation is applied involves generally different  states $|\Phib^{(2)}_k\rangle$,      one cannot simply  project      the relative   permutation on $\Mc_{2n}$,   as was done in section \ref{sec3B}.     In this case one can  reverse the substitution  of the symmetrization projectors by the matching operators, performed in section \ref{sec3}, i.e.,   replace back  
\be
\hat{\Mc}_{2n } \to \hat{S}_{2n}
\en{PtoS}
in   Eq.   (\ref{E17d}). The matchings are then  replaced by permutations and the normalization factor is   adjusted accordingly. In this way one can get   the output probability in the  form 
\begin{eqnarray} 
 \label{E17new}
&&p_\mb =    \frac{1}{\mb!} \frac{p_\mathbf{0}}{\left(2^nn!\right)^2 } \sum_{k_1=1}^N \ldots \sum_{k_n=1}^N  \sum_{k^\prime_1=1}^N \ldots \sum_{k^\prime_n=1}^N \sum_{\sigma\in S_{2n}} \sum_{\tau\in S_{2n}} \nonumber\\
&&  \!\!\!  \times \left[ \prod_{i=1}^n  r_{k_i}  \U^*_{k_i ,{\sigma(2i-1)}}\U^*_{k_i ,{ \sigma(2i)}} r_{k^\prime_i}  \U_{ k^\prime_i,{\tau(2i-1)}} 
\U_{k^\prime_i,{\tau(2i)}} \right]\nonumber\\
    && \times \langle \Phib^{(2)}_{k_1}|  \ldots  \langle \Phib^{(2)}_{k_n}| \hat{P}_{\sigma^{-1}\tau} |\Phib^{(2)}_{k^\prime_1}\rangle \ldots |\Phib^{(2)}_{k^\prime_n}\rangle. \nonumber\\
  \end{eqnarray}
 In this most general case,   the   distinguishability  function  is defined by a  complicated expression involving  the sets  of  input indices $k_1,\ldots, k_n$ and $k^\prime_1, \ldots, k^\prime_n$ as well as     permutation $\sigma\in S_{2n}$:
\be
J_{\mathbf{k}, \mathbf{k}^\prime}(\sigma) =   \langle \Phib^{(2)}_{k_1}|  \ldots  \langle \Phib^{(2)}_{k_n}| \hat{P}_{\sigma} |\Phib^{(2)}_{k^\prime_1}\rangle \ldots |\Phib^{(2)}_{k^\prime_n}\rangle.
\en{JGen}
There is  no  other  symmetry   in Eq. (\ref{JGen}) apart from the fact that permutations of the photons within each pair do not change the   internal states (in the degenerate case). Thus one can reduce the permutation group $S_{2n}$ in Eq. (\ref{JGen}) to the factor group $S_{2n}/S^{\otimes n}_2$. 

%%%%%%%%%%%%%%%%%%%%%%%%%%%%%%%%%%%%%%%%%%%%%%%%
\section{ Measure of  indistinguishability } 
\label{sec4}

Let us  further analyze the effect of distinguishability in  the case of identical internal states $|\Phib^{(2)}_k\rangle  = |\Phib^{(2)}\rangle$ considered in section \ref{sec3B}.  Our goal  is to quantify the indistinguishability of photons in  interference with such   squeezed states.  We focus on the case of  the  degenerate squeezed states at interferometer input.    The output probability of Eq. (\ref{E17dB}) can  be written also  in a similar form as in  Eq. (\ref{E32}) of section \ref{sec3C}, 
 \begin{eqnarray} 
 \label{E32A}
&&    p_\mb =    \mathrm{Tr}\left\{ \hat{\Pi}_\mb   \prod_{k=1}^N|r_k)  (r_k|  \right\} =   p_\mathbf{0}\binom{2n}{n}\frac{ (2n)!}{\mb!}  \nonumber\\ 
&&  \times \langle \Psib^{(2)}|^{\otimes n}   \hat{\Mc}^\dag_{2n}  \left[\prod_{i=1}^{2n} |l^{(out)}_i\rangle\langle l^{(out)}_i| \otimes  \openone  \right]    \hat{\Mc}_{2n}   |\Psib^{(2)}\rangle^{\otimes n}  \nonumber\\
\end{eqnarray}
where  
\begin{eqnarray}
\label{E35}
|\Psib^{(2)}\rangle&= &\left[\sum_{k=1}^N  \frac{r_k}{2} |k \rangle |k\rangle \right] \otimes |\Phib^{(2)}\rangle \\
&=&\frac12\left[\sum_{l=1}^M \sum_{s=1}^M A_{ls} |l^{(out)}\rangle| s^{(out)} \rangle\right]\otimes |\Phib^{(2)}\rangle, \nonumber
\end{eqnarray}
with $A_{ls} = \sum_{k=1}^N U_{kl}r_kU_{ks}$. 

%%%%%%%%%%%%
\subsubsection{A  measure of indistinguishability    }
\label{secq2n}
Let us  decompose the internal  state   $|\Phib^{(2)}\rangle^{\otimes n}$ of $2n$ photons into the symmetric part 
 and an orthogonal  complement, $ |\Phib^{(2)}\rangle^{\otimes n} =  \hat{S}_{2n} |\Phib^{(2)}\rangle^{\otimes n} + (\openone -  \hat{S}_{2n}) |\Phib^{(2)}\rangle^{\otimes n}$.    We will use the  following factorization  identity  
\be
\hat{\Mc}_{2n} \left(\openone \otimes \hat{S}^{(int)}_{2n}\right) = \hat{\Mc}^{(op)}_{2n} \otimes \hat{S}^{(int)}_{2n},
\en{E34}   
where the operator $ \hat{\Mc}_{2n}^{(op)}$   acts on the operational   modes only. Eq. (\ref{E34})  can be easily established using the operator  composition rule
\[
\left(\hat{P}_\alphb \otimes \hat{P}_\alphb\right)\left(\openone\otimes \hat{P}_\sigma\right) = \hat{P}_\alphb \otimes \hat{P}_{\alphb\sigma}
\]
and observing that   the permutation   $\tau\equiv \alphb\sigma$  enumerates all elements of   $S_{2n}$.    The symmetric part of the internal state $ \hat{S}_{2n} |\Phib^{(2)}\rangle^{\otimes n}$  corresponds to    the completely indistinguishable case, since  such an  internal state factors out, due to the identity in Eq. (\ref{E34}),  and does not contribute to the output probability in  Eq. (\ref{E32A}).   
 
Let $ q_{2n} $ be the    probability that the    internal state  of $2n$ photons  is symmetric,  i.e., 
\begin{eqnarray} 
 \label{q2n}
 q_{2n} & \equiv  &   \langle\Phib^{(2)}|^{\otimes n} \hat{S}_{2n}|\Phib^{(2)}\rangle^{\otimes n}  =  \langle\Phib^{(2)}|^{\otimes n} \hat{\Mc}_{2n}|\Phib^{(2)}\rangle^{\otimes n} \nonumber\\
 & = & \frac{1}{(2n-1)!!} \sum_{\alphb\in \Mc_{2n}} J(\alphb).
 \end{eqnarray}
 Though below we discuss   the probability $q_{2n}$ only in the case of identical internal states of photon pairs, this probability  can be defined   in the general case of different internal states
\be 
 q_{2n} =    \langle\Phib^{(2)}_{k_1}|\ldots \langle\Phib^{(2)}_{k_n}|   \hat{S}_{2n}  |\Phib^{(2)}_{k^\prime_1}\rangle\ldots  |\Phib^{(2)}_{k^\prime_n}\rangle,
\en{q2nGen} 
with the  explicit dependence on  the input ports of the considered  photon  pairs  (reflected by the indices $k_j$ and $k^\prime_j$).

The   introduced probability  $q_{2n}$ is the probability that  $2n$  photons are   indistinguishable, quite   similarly as in  the case of  single photons at interferometer input   \cite{TBonBS}.    When  $q_{2n}=1$  the  identity (\ref{E34}) implies that the  photons  interfere  as  completely  indistinguishable, i.e.,  a completely  symmetric internal state of $n$ photon pairs  has  no influence on the output probability distribution.  Such a symmetry corresponds to  single-mode  squeezed states with the same internal mode for all photons.        

For identical  internals states of photon pairs  the probability $q_{2n}$ in Eq. (\ref{q2n}) satisfies    $q_{2n}\ge \frac{1}{(2n-1)!!}$  thanks to  the  symmetry of the tensor product of   internal states  under the  permutations of   photon pairs and transpositions of two photons in each photon pair. The lower bound is  the  lowest possible indistinguishability    due to the  coinciding  internal states. It can be shown (see appendix \ref{appD} for details)
 that   the probability $q_{2n}$  in  Eq. (\ref{q2n})  can be cast as 
\begin{eqnarray} 
 \label{E37}
q_{2n}    &=&  \sum_{|\bs|=n}\binom{n}{\bs}   \left(\prod_{j=1}^\infty p_j^{s_j} \right)\frac{\prod\limits_{j=1}^\infty(2s_j-1)!!}{(2n-1)!!}\\
&=& \binom{2n}{n}^{-1}\sum_{|\bs|=n}  \prod_{j=1}^\infty \binom{2s_j}{s_j} p_j^{s_j},\nonumber
 \end{eqnarray}  
 where  $\bs = (s_1,s_2,\ldots)$, with $s_j$ being the   number of occurrences of  the internal mode  $|\phi_j\rangle$ in the tensor product of the internal states $|\Phib^{(2)}\rangle^{\otimes n}$, when the latter is expanded over the tensor products of the internal modes.

The first  expression in  Eq. (\ref{E37})   confirms our interpretation of $q_{2n}$ as   the probability of photons behaving as indistinguishable:   the multinomial distribution  $\binom{n}{\bs} \prod_j p_j^{s_j}$ gives the probability of a particular subset  of the internal  modes  $|\phi_j\rangle$  of  $2n$ interfering photons, whereas the last  factor is the probability that in each   matching  pair the  photons  have the same internal states  (only  the indistinguishable photons   interfere). 

For instance, if there are only two detected photons, they come   from the same (degenerate) squeezed state, hence they are indistinguishable. We have $q_2 = 1$.  For  four detected photons,   there are two combinations of non-zero  occupations $\bs = (s_1,s_2,\ldots)$  of photons in the internal modes: $s_j = 2$   and $s_{j_1}= s_{j_2}=1$. Hence, we obtain from Eq. (\ref{E37}): 
\begin{eqnarray}
q_4 &=& \binom{4}{2}^{\!-1} \left[\sum_{j=1}^\infty \binom{4}{2} p^2_j +  \sum_{j_1< j_2} \binom{2}{1}^{\!2} p_{j_1}p_{j_2} \right]\nonumber\\
&=& \frac{1+2 \mathbb{P}}{3},
\label{q_4}\end{eqnarray}
where we have used the identity \mbox{$\sum_{j_1< j_2} p_{j_1}p_{j_2}  = \frac12(1-\sum_j p^2_j)$} and the definition of the purity   Eq. (\ref{SchmidtK}).    For  $n\ge 3$,  Eqs. (\ref{E25}) and (\ref{q2n}) indicate that that  the probability $q_{2n}$   depends also   on the higher-order moments of the singular values $ \sum_j p^s_j$, $s\le  n$.

 The lowest possible  indistinguishability $q_{2n} = \frac{1}{(2n-1)!!}$   is attained in Eq. (\ref{E37}) for  divergent  Schmidt number   in  Eq. (\ref{SchmidtK}) $K \to \infty$ (or, equivalently,  for vanishing purity $ \mathbb{P}\to 0$), i.e., when $p_j\to 0$  whereas $\sum_j p_j =1$. In this limit all  higher moments  of the singular values, starting from the purity, vanish, $\sum_j p^s_j\to 0$. By   using the relation between summations  Eq. (\ref{E5}) of section \ref{sec2}, we obtain in this case from    Eq. (\ref{E37})  
   \be
 \lim_{\mathbb{P}\to 0} q_{2n}= \frac{1}{(2n-1)!!}\sum_{j_1=1}^\infty \ldots \sum_{j_n=1}^\infty \prod_{i=1}^n p_{j_i}  = \frac{1}{(2n-1)!!},
 \en{q2nLim}
 where  we have taken into account that the  coincidences $j_i = j_{i^\prime}$ give a vanishing contribution.

%%%%%%%%%%%%
\subsubsection{Bound on the total variation distance    }
\label{secVDb}
 
From   Eqs. (\ref{E32A})-(\ref{q2n})     we  obtain the decomposition of the output probability   $p_\mb$  as follows
 \be
p_\mb = q_{2n}  \mathring{p}_\mb  + (1-q_{2n}) p^{(\perp)}_\mb, 
\en{E36} 
 where $ \mathring{p}_\mb$ is  given by Eq. (\ref{E18}), whereas  the complementary  probability     $p^{\perp}_\mb$   is obtained by  replacing the internal state of $2n$ photons  by  the complementary part,    orthogonal   to the symmetric subspace, 
 \[
 |\Phib^{(2)}\rangle^{\otimes n} \to \frac{  \openone- \hat{S}_{2n}}{\sqrt{1-q_{2n}}} |\Phib^{(2)}\rangle^{\otimes n} 
 \]
 i.e.,  the input state of Eq. (\ref{E35})  is replaced with the following one 
   \begin{eqnarray}
   |\Psib^{(2n)}_\perp\rangle \equiv  \openone\otimes \frac{  \openone- \hat{S}^{(int)}_{2n}}{\sqrt{1-q_{2n}}} |\Psib^{(2)}\rangle^{\otimes n},
\end{eqnarray} 
where  $  |\Psib^{(2)}\rangle$ is the state in Eq. (\ref{E35}). Observe that by construction  $p^{(\perp)}_\mb$  is  a normalized  probability distribution
\begin{eqnarray}
\label{normp}
   \sum_{n=0}^\infty\sum_{|\mb|=2n} p^{(\perp)}_\mb =  1.
\end{eqnarray}

Consider  now  the total variation distance  between the  output probability distribution $p_\mb$ of  Eq. (\ref{E36})    and  that of  the ideal case $ \mathring{p}_\mb$ Eq. (\ref{E18}).  From Eq. (\ref{E36}) we obtain: 
\begin{eqnarray}
\label{E39}
\mathcal{D} &\equiv& \frac12\sum_{n=0}^\infty  \sum_{|\mb|=2n} |\mathring{p}_\mb - p_\mb | \nonumber\\
&=& \frac12\sum_{n=0}^\infty  (1-q_{2n})\sum_{|\mb|=2n} |\mathring{p}_\mb - p^{(\perp)}_\mb |\nonumber\\
& \le&   \sum_{n=0}^\infty (1-q_{2n}) \sum_{|\mb|=2n}  \mathring{p}_\mb = 1 - \bar{q}.
\end{eqnarray}  
Here   we have used that the variation distance is bounded by the total probability  
\[
\frac12 \sum_{|\mb|=2n} |\mathring{p}_\mb - p^{(\perp)}_\mb |\le \sum_{|\mb|=2n} \mathring{p}_\mb  \equiv  \mathring{p}(2n)
\]
and   introduced the averaged probability $\bar{q}$, where the averaging is over the ideal distribution
$ \mathring{p}_\mb$,  
\begin{eqnarray}
\label{E40}
  \bar{q} \equiv    \sum_{n=0}^\infty   \mathring{p}(2n)  q_{2n}.
\end{eqnarray}
We  have      (see appendix \ref{appE})
\begin{eqnarray}
\label{E41}
   \mathring{p}(2n) =  \mathring{p}_{\mathbf{0}}\sum_{|\nb|=n} \prod_{k=1}^N\binom{2n_k}{n_k}\left(\frac{r_k}{2}\right)^{2n_k} 
\end{eqnarray}
with $  \mathring{p}_{\mathbf{0}}=  \mathring{\Zc} =   \prod\limits_{k=1}^N\  \left(1- r_k^2 \right)^{\frac14}$.  

 Eqs. (\ref{q2n}), (\ref{E37}), (\ref{E36}),  and (\ref{E39}) allow one to  interpret $\bar{q}$ as the  measure of  average indistinguishability, an analog of a similar measure  in the case of interference with single photons at   interferometer input  \cite{TBonBS}. 

%%%%%%%%%%%%%%%%%%%%%%%%%%%%
\subsection{ Estimate of indistinguishability in the  Gaussian boson sampling experiment}

We will use that for    $N$ equally  squeezed single-mode states,   $r_k = r$ for $k=1,\ldots, N$,   the probability to detect $2n$ photons has a simple form, reminiscent  of the negative binomial distribution   (with  a half-integer  number of successes  $N/2$, in general; see appendix \ref{appE})
\be
 \mathring{p}(2n)  =  (1-r^2)^{\frac{N}{2}} \left(\frac{N}{2}\right)_n \frac{r^{2n}}{n!},
\en{E42}
where $(m)_n = m(m+1) \ldots (m+n-1)$. With simple algebra one can get  the average number of photon pairs and the relative dispersion
\be
\bar{n}  = \frac{N}{2}\frac{r^2}{1-r^2},\quad  \frac{\overline{n^2} - \bar{n}^2}{\bar{n}^2}  = \frac{2}{Nr^2}.
\en{nbar}
Observe that by Eq. (\ref{nbar})  for  $N\gg1/r^2$ the  distribution of the number of detected photons,  Eq. (\ref{E42}), becomes sharp about the average,  allowing  to  approximate   the average probability of indistinguishable photons $\bar{q}$  Eq. (\ref{E40}) by   the most probable value 
  $\bar{q} \approx q_{2\bar{n}}$. The latter  can be applied also for  the case of almost  equal  squeezing parameters $r_k\approx r$ (recall that      
  $r = \tanh\kappa$, where  $\kappa$  is usually called the squeezing parameter). 
  
 In the case when the squeezed states are   very close to being the single-mode states,  one can employ  the two-mode approximation consisting of   the most probable mode and   noise (such a model was used in Ref. \cite{ExpGBS2} to characterize purity of the squeezed states). Let      the two singular values be   $p_1 = 1-\epsilon$ and $p_2 = \epsilon$, for  some $\epsilon \ll 1$.  The noise  amplitude   $\epsilon$  is related to the  purity
\be
\mathbb{P} \equiv \sum_{j=1,2} p^2_j = (1-\epsilon)^2 +\epsilon^2.
\en{E44} 
The two-mode approximation allows to easily   evaluate the sum in Eq. (\ref{E37}) and estimate the value of  $q_{2\bar{n}}$. From  the second expression in  Eq. (\ref{E37}) we obtain 
\begin{eqnarray}
 \label{E45}
&& \bar{q}\approx q_{2\bar{n}} =  \binom{2\bar{n}}{\bar{n}}^{\!-1}\!\!\sum_{s=0}^{\bar{n}}  \binom{2s}{s}  \binom{2\bar{n}-2s}{\bar{n}-s}   \epsilon^s (1-\epsilon)^{\bar{n}-s}
\nonumber\\
&& =(1-\epsilon)^{\bar{n}}\mathcal{F}\left(\frac12,-\bar{n},\frac12-\bar{n}, \frac{\epsilon}{1-\epsilon}\right) \approx    (1-\epsilon)^{\bar{n}},
\end{eqnarray}
where $\mathcal{F}$ is the Gauss Hypergeometric function,   thus  $ \mathcal{F}(1/2,-\bar{n},1/2-\bar{n},x) \approx  1$ for $0< x\ll 1$ and arbitrary  $\bar{n}$. 
Eq. (\ref{E45}) predicts  that the probability of photons behaving as indistinguishable  in an interference with imperfectly single-mode squeezed states  falls exponentially fast  in the average number of detected photons. 

The   recent  experimental  Gaussian  boson sampling    \cite{ExpGBS2}   corresponds to    $N = 50$ degenerate squeezed vacuum states    (in an equivalent representation of $N=25$ non-degenerate input states in $N = 50$ input ports, see section \ref{sec3})   and  the squeezing parameters  $r_k\sim 1$.  Let us estimate the indistinguishability in this experiment by employing the  average case approximation $\bar{q}\approx q_{2\bar{n}}$ and  the above two-mode  model  of   noise.  The average  reported experimental purity in Ref.    \cite{ExpGBS2}   is    $\overline{\mathbb{P}} = 0.938$, hence  $\epsilon = 0.032$ by Eq. (\ref{E44}).    The average number of detected photons  $2\bar{n} \ge  43$, for   the average number of clicks  is  reported to be  43.    Therefore,  by Eq. (\ref{E45})   the  average  probability  of photons behaving as completely  indistinguishable satisfies    $q_{2\bar{n}} \le 0.5$.

The above discussion leaves out one important question:  How can one estimate the indistinguishability parameter  $q_{2n}$ from an experiment?
  Can one  estimate the average  indistinguishability  $q_{2\bar{n}}$  directly from the limited  experimental data obtained  in experiments on the   Gaussian  boson sampling?  Since limited data allow only to estimate some  low-order correlations,  is it possible to estimate  this parameter from such  low-order correlations, e.g., by considering the correlations in a few  output ports?     The following point should be taken into account:  if     less than $2\bar{n}$  photons are detected,  no higher-order cycles  contributing  to $q_{2\bar{n}}$ (see  Eqs. (\ref{E25}) and (\ref{q2n}) and  also appendix \ref{appD}) can influence the experimental data.  This fact does not allow  to directly estimate $q_{2\bar{n}}$ from the low-order correlations, since the latter correspond to  much smaller  photon numbers as  compared to the average total  number of detected photons.       A similar problem arises also   in the  case of   interference and the boson sampling with  single photons, see Refs. \cite{SYMDTh,QPaper}. For instance, in Ref. \cite{QPaper} it was shown that the  low-order correlations  would be insufficient to distinguish such boson sampling from efficient  classical  approximations. Similarly here,    direct estimate of $q_{2\bar{n}}$  from an experiment requires going beyond the low-order correlations.  One way out would be to estimate the purity by the four-photon detection  in an interference on a beamsplitter  (by using  pairs of the degenerate squeezed states or a single non-degenerate squeezed state at a time).  From Ref. \cite{ModeStrInfl} and also from  Eq. (\ref{BSprob}) of section \ref{sec3B}     it is  seen  that   the output probability depends on the purity. After that one can get an estimate on the indistinguishability using the above two-mode model.

%%%%%%%%%%%%%%%%%%%%%%%%%%%%%%%%%%%%%%%%%%%%%%%% 
\section{ Non-Gaussian  squeezed states  } 
\label{sec5}

In the previous sections we have seen the power of the first-order quantization representation for analysis of quantum interference with the Gaussian  squeezed states. The  purpose of this section is to investigate how  the approach can be extended     to   generalized (non-Gaussian) squeezed vacuum states  \cite{GenSQ}. Such squeezed states are   produced by $\mu$-photon processes with $\mu\ge 3$, such as in the recent experimental demonstration of the three-photon spontaneous parametric down-conversion   \cite{3phDC}.   In   the  parametric approximation, the     multi-mode  generalized squeezed state  can be represented by the following exponential operator 
\be 
|A) \propto  \exp\left\{\sum_{i_1\in I_1}\ldots  \sum_{i_\mu\in I_\mu} A_{i_1\ldots i_\mu} \hat{c}^\dag_{i_1} \ldots \hat{c}^\dag_{i_\mu} \right\}|0).
\en{E51}
The exponent in Eq. (\ref{E51})  has  divergent  power series expansion  for $\mu>2$     \cite{Hillery}, as the    parametric approximation  disregards   power depletion  in the  optical pump  \cite{SPent,EnConDC}.   However, for  a finite total number of detected  photons  one  needs to retain  only some  finite number of terms of the divergent Taylor series.      Below  the focus will be  on the degenerate case   corresponding  to  \mbox{$I\equiv I_1=I_2=\ldots =I_\mu$} and   a symmetric  tensor $A$.    

For  the  Gaussian  squeezed states,  $\mu=2$,   the existence of  the singular value decomposition of matrices allows one to            diagonalize the complex symmetric  (generally, infinite-dimensional) matrix $A$ in  Eq. (\ref{E51}) to the Schmidt modes  \cite{MMSS}. There is  a unitary  (also, in general,  infinite-dimensional) matrix    $V_{ij}$ that
\be
 A_{il} = \sum_{j\in I} \lambda_jV_{ij} V_{lj},
\en{E52}
with $\lambda_j\ge 0$  being  the singular values and columns of $V$ the Schmidt modes $\phi_j$. Introducing new boson creation operators  by the same unitary transformation 
 \be
  \hat{a}^\dag_j = \sum_{i\in I} \hat{c}^\dag_i V_{ij}
  \en{E53}
we get the diagonal form of the multi-mode Gaussian squeezed states, which was the starting point  in section \ref{sec2},  where  $\lambda_j = r\sqrt{p_j}$.    
 
 For  $\mu\ge 3$ the   symmetric tensor $A$ in Eq. (\ref{E51})  can also be similarly diagonalized as a convex  sum of the  tensor-products of vectors   \cite{TensorDiag}  (the columns of the matrix $V$)
 \be
 A_{i_1\ldots i_\mu} = \sum_{j\in I} \lambda_j V_{i_1j} \ldots V_{i_\mu j}.
 \en{E54}
However,  in this case one has to use, in general,  a  non-unitary  matrix   $V_{ij}$. 
Nevertheless, we   can introduce  new boson creation operators, similarly   to  Eq. (\ref{E53}),   in order  to diagonalize the expression in the  exponent  in Eq. (\ref{E51}), even if  they  correspond  to some  non-orthogonal states. Indeed,  the   new boson  creation operators   can be used   in the   identity  Eq. (\ref{E3}), relating the first-order and second-order quantization representations,  as the latter  remains valid irrespective orthogonality of the single-particle states. 

Thus the  generalized squeezed states  can be  reduced  in  the first-order quantization representation   to a simpler  diagonal form, similarly as the Gaussian squeezed states.  Consider  $N$   multi-mode  squeezed $\mu$-photon  states  with the overall  squeezing parameters $r_1,\ldots, r_N$ (introduced similarly as in section \ref{sec2}, by rescaling $\lambda_j = rp_j^{1/\mu}$ of the singular values in Eq. (\ref{E54}), where the positive parameters $p_j$ sum to $1$). Similarly as in section \ref{sec3}, the combined state of $N$ generalized squeezed  states can be cast as follows:
\begin{eqnarray}
\label{E55}
&& \prod_{k=1}^N |r_k) \propto \sum_{n=0}^\infty \frac{\sqrt{(\mu n)!}}{n!} \hat{S}_{\mu n} \left[ \sum_{k=1}^N r_k \otimes |\Phib^{(\mu)}_k\rangle\right]^{\otimes n},\qquad
 \\
&& |\Phib^{(\mu)}_k\rangle \equiv \sum_{j\in I} \left(p^{(k)}_j\right)^{\frac{1}{\mu}}|\phi^{(k)}_j\rangle^{\otimes \mu}. \nonumber
\end{eqnarray}
Now, the state in Eq. (\ref{E55}), to which the projector $\hat{S}_{\mu n}$ is applied, is symmetric with respect to permutations of $\mu$-tuples of photons   and with respect to  permutations of the  photons in each $\mu$-tuple.  Let  $\Mc^{(\mu)}_{\mu n}$ be  the set of all $\mu$-dimensional matchings, i.e.,   partitions of $\mu n$ elements $\{1,\ldots, \mu n\}$ into $n$      disjoint $\mu$-tuples $(\alpha_{i\mu+1},\ldots,\alpha_{(i+1)\mu})$,  $i=1,\ldots, n$,  where  permutations  of the  elements in each $\mu$-tuple do not produce new partitions.   The set   $\Mc^{(\mu)}_{\mu n}$ can be enumerated by a vector-index  $ \alphb^{(\mu)}\equiv (\alpha_1,\ldots,\alpha_{\mu n})$,  if we order the $\mu$-dimensional matchings by the first element, where in each $\mu$-tuple we choose as the first element   the smallest one by  permutation of the elements. It is easy to establish that  there are
\be  
 (\mu n-1)!^{(\mu)} \equiv  \frac{(\mu n)!}{(\mu!)^n n!} 
 \en{E56} 
 $\mu$-dimensional matchings in $\Mc^{(\mu)}_{\mu n}$. We can project an arbitrary permutation   $\sigma\in S_{\mu n}$    on $\Mc^{(\mu)}_{\mu n}$ by expanding  $\sigma$ as follows
\be
\sigma =   \pi \left( \tau_1\otimes \ldots \otimes \tau_n \right)   \alphb,
\en{E57} 
where  $\pi \in S_n$ permutes  $n$ $\mu$-tuples, and $\tau_i  \in S_{\mu}$ permutes  the   elements of  the $i$th  $\mu$-tuple.    The symmetrization projector  $\hat{S}_{\mu n}$  can be factored accordingly  
\begin{eqnarray}
\label{E58}
&& \hat{S}_{\mu n} =    \hat{S}^{(tuple)}_n\hat{S}_\mu^{\otimes n}\hat{\mathcal{M}}^{(\mu)}_{\mu n} ,  \nonumber\\
&&  \hat{\mathcal{M}}^{(\mu)}_{\mu n} \equiv \frac{1}{(\mu n-1)!^{(\mu)}}\sum_{\alphb\in \Mc^{(\mu)}_{\mu n}} \hat{P}_\alphb.
 \end{eqnarray}
Now, due to the symmetry by construction of the state in Eq. (\ref{E55}),    the $\mu$-dimensional matching operator $\hat{\mathcal{M}}^{(\mu)}_{\mu n}$ can replace the projector $\hat{S}_{\mu n}$, quite similarly as in the case of the Gaussian  squeezed states. One can then proceed from this point.

Summarizing the above,     the first-order quantization representation is  suitable  also or the generalized squeezed states, with, however,  a new feature:  the equivalent of the Schmidt modes in the diagonal representation are not mutually orthogonal, in general.   

%%%%%%%%%%%%%%%%%%%%
\section{Conclusion}
\label{sec6}

In conclusion, the  first-order quantization representation, commonly  underestimated,    proves to be extremely  useful  approach to study the  quantum interference  of  the squeezed  vacuum  states on a unitary interferometer. It  allows  for straightforward  derivation of the output probability distribution with  account the fact that   realistic squeezed states  possess  continuous degrees of freedom,  called the  Schmidt modes.   The method  also reproduces   previously known results in the limiting cases, e.g.,  it reproduces the probabilities for the  four-photon interference  on a beamsplitter and  the well-known  probability formula for   the  case  of the  squeezed states   in a single common  Schmidt mode.

 It is found that  the  multi-mode structure (i.e., several Schmidt modes)  is  one of the sources of    distinguishability of  the squeezed states:  each photon  pair is effectively in a  mixed internal state, which leads to partial distinguishability.     A  quantitative measure of indistinguishability, $q_{2n}$  is proposed. It  is the probability that $n$ pairs of  photons interfere as indistinguishable. Moreover,  it   bounds the total variation distance  to the  output distribution of the ideal  indistinguishable case.  In this respect, the proposed measure  of indistinguishability is quite similar to that     for single photons.  It is shown that    $q_{2n}$     decreases   exponentially fast  in $n$. For example, the  recent   Gaussian boson sampling  experiment    with  the reported  purity $\mathbb{P}\approx 0.938$  is, on average,    close to  the middle line between distinguishable and indistinguishable cases with    $q_{2\bar{n}}\lesssim  0.5$ for  $2\bar{n}\ge 43$.  This  fact  apparently means that   partial distinguishability   has also  a  strong effect on the  computational complexity of the output probability distribution from  an experimental Gaussian boson sampling.   It is known that     distinguishability  of single photons has a strong effect on the computational complexity of the usual  boson sampling.
   
Finally, the approach of this work  is not limited only  to the Gaussian states, as it allows for  generalization to the  generalized (non-Gaussian) squeezed states. Such generalized squeezed states were  already observed   in the recent three-photon down conversion experiment. 
 
\section{acknowledgements}
The author  is grateful  to anonymous referees whose detailed comments helped to improve the presentation.  This work  was supported by the National Council for Scientific and Technological Development (CNPq) of Brazil,  Grant 307813/2019-3.

\bigskip
\appendix

%%%%%%%%%%%%%%%%%%%%%%%%%%%%%%%%%%%%%%%%%%%%
 \section{ Matchings  do  not  form  a group } 
\label{appB}

Counterexamples to the group properties are easily found  for   $n\ge 3$. Consider the  following matching permutation $\alphb =  (1,5,2,4,3,6)\in \Mc_{6}$.  We have  $\alphb^{-1}(3) = 5 > \alphb^{-1}(4) = 4$, hence $\alphb^{-1}\notin \Mc_{6}$.  Additionally,   $\alphb^2 = (1,3,5,4,2,6)$ with   $\alphb^2(3) = 5 > \alphb^2(4) = 4$, hence  also  $\alphb^2 \notin \Mc_{6}$.  The action of $\alphb$ is further illustrated in Eq. (\ref{B1}):
\begin{eqnarray}
\label{B1}
\begin{array}{cccccc} \alphb(1) = \alpha_1 = 1, \\  \alphb(2) = \alpha_2 = 5, \\ \alphb(3) = \alpha_3 = 2, \\ \alphb(4) = \alpha_4 = 4, \\ \alphb(5) = \alpha_5 = 3, \\ \alphb(6) = \alpha_6 = 6,
 \end{array} \qquad \begin{array}{cccccc} \alphb^2(1) = \alphb(\alpha_1) = 1, \\  \alphb^2(2) = \alphb(\alpha_2) = 3, \\ \alphb^2(3) = \alphb(\alpha_3) = 5, \\ \alphb^2(4) = \alphb(\alpha_4) = 4, \\ \alphb^2(5) = \alphb(\alpha_5) = 2, \\ \alphb^2(6) = \alphb(\alpha_6) = 6.
 \end{array}
\end{eqnarray} 

%%%%%%%%%%%%%%%%%%%%%%%%%%%%%%%%%%%%%%%%%%%%
\section{Derivation  of the  output  probability } 
\label{appA}

Substituting Eqs. (\ref{E8}), (\ref{Proj}) and (\ref{E15}) of the main text  in  the Born rule and using   the identity $(2n)!\binom{2n}{n} \frac{1}{\left[(2n-1)!!\right]^2} = 2^{2n}$  we obtain
\begin{eqnarray}
 \label{A1}
 && p_\mb =  \mathrm{Tr}\left\{ \hat{\Pi}_\mb   \prod_{k=1}^N|r_k)  (r_k|  \right\}  \\
 && = \frac{ \Zc^2}{\mb!}   \sum_{\alphb,\betb} \left[ \sum_{k=1}^N r_k \langle k|\langle k|  \otimes \langle \Phib^{(2)}_k|  \right]^{ \!\otimes n}  \hat{P}^\dag_\alphb \otimes \hat{P}^\dag_\alphb
 \nonumber\\
 && \times    \left[\prod_{i=1}^{2n}  |l^{(out)}_{i}\rangle\langle l^{(out)}_{ i}| \otimes  \openone \right]   \hat{P}_\betb \otimes  \hat{P}_\betb   \nonumber\\
 && \times\left[\sum_{k=1}^N r_k |k\rangle|k\rangle \otimes | \Phib^{(2)}_k\rangle \right]^{ \!\otimes n}.\nonumber
\end{eqnarray}
Now we  apply the matching operators $\hat{P}_\alphb$ and $\hat{P}_\betb$ to the output states and   expand the tensor products of    the linear combinations of two-photon   input states. With the use of the   input-output relation Eq. (\ref{E10}) we get
\begin{eqnarray} 
 \label{A3}
&& p_\mb =  \frac{ \Zc^2}{\mb!} \sum_{\alphb,\betb} \sum_{k_1=1}^N \ldots \sum_{k_N=1}^N  \sum_{k^\prime_1=1}^N \ldots \sum_{k^\prime_N=1}^N \nonumber\\
&& \times \left[ \prod_{i=1}^n  r_{k_i}  U^*_{k_i l_{\alpha_{2i-1}}} U^*_{k_i l_{\alpha_{2i}}}  r_{k^\prime_i}  U_{ k^\prime_i l_{\beta_{2i-1}}} 
U_{k^\prime_i l_{\beta_{2i}}} \right]\nonumber\\
    && \times \langle \Phib^{(2)}_{k_1}|  \ldots  \langle \Phib^{(2)}_{k_N}| \hat{P}^\dag_{\alphb}\hat{P}^{}_{\betb}|\Phib^{(2)}_{k^\prime_1}\rangle \ldots |\Phib^{(2)}_{k^\prime_N}\rangle,
\end{eqnarray}
which is  Eq. (\ref{E17d}), taking  into account Eq. (\ref{MU}). 

%%%%%%%%%%%%%%%%%%%%%%%%%%%%%%%
  \section{ Cycle index  over matchings } 
\label{appC}
Recall that for a permutation group $S_n$ acting on the set $\{1,2,\ldots, n\}$, the cycle index   \cite{Stanley}   is  the sum  
\begin{eqnarray}
\label{C1}
 \Theta_n  &\equiv& \sum_{\sigma\in S_n} t_1^{C_1(\sigma)} t_2^{C_2(\sigma)}\ldots t_n^{C_n(\sigma)} \nonumber\\
& = &\sum_{C_1,\ldots,C_n} \#_{S_n}(C_1,\ldots,C_n) t_1^{C_1} t_2^{C_2 }\ldots t_n^{C_n},  \nonumber \\
&&  \#_{S_n}(C_1,\ldots,C_n)\equiv  \frac{n!}{\prod_{k=1}^n k^{C_k}C_k!},
\end{eqnarray}
where   $t_1,\ldots, t_n$ are free  parameters, $C_k$ is the number of $k$-cycles in the   disjoint cycle decomposition of a permutation, the  sum over $C_1,\ldots, C_n$ is  conditioned on  $\sum_{k=1}^n kC_k = n$, and  the factor  $\#_{S_n}(C_1,\ldots,C_n)$ is equal to the total number of permutations  $\sigma\in S_n$ with a given  cycle structure  $(C_1,\ldots, C_n)$.

Let us now  consider a similar  cycle index but on the  cycles  of   matching permutations $\alphb\in \Mc_{2n}$ Eq. (\ref{Vind}) acting on the double-set $(1,1,2,2,\ldots, n,n)$, i.e., 
\begin{eqnarray}
\label{C2}
\Xi_n &\equiv&  \sum_{\alphb\in \Mc_{2n}}  t_1^{C_1(\alphb)} t_2^{C_2(\alphb)}\ldots t_n^{C_n(\alphb)} \\
&=& \sum_{C_1,\ldots,C_n} \#_{\Mc_{2n}}(C_1,\ldots,C_n) t_1^{C_1} t_2^{C_2 }\ldots t_n^{C_n},\nonumber
\end{eqnarray}
 where the  sum over $C_1,\ldots, C_n$ is  conditioned on  $\sum_{k=1}^n kC_k = n$, and the cycle decomposition of  a matching, when acting on a double-set, is  defined in section \ref{sec3B}.  
 
 We need to count  the number of  matchings  $\#_{\Mc_{2n}}(C_1,\ldots,C_n)$, from the total number  $(2n-1)!!$, which have a given cycle structure  $(C_1,\ldots, C_n)$.    Consider a   matching permutation of some  given cycle structure, say   $\alphb\equiv  \nu_1\nu_2\ldots\nu_q \in (C_1,\ldots, C_n)$, where $\nu_i$ are the  cycles as  defined by  Eq. (\ref{Cycle}) of section \ref{sec3B}. To count the number of matchings for a given cycle structure  it is convenient to  convert the cycles over the double-set of $2n$ elements $1\le x_i\le  n$ (i.e., with the elements repeated twice) to similar  matching cycles over a set of $2n$ distinct elements. This can be done  by adding   the number  $n$ to the second  element   in each pair, e.g.,  we make the following transformation of  a $k$-cycle defined by  Eq. (\ref{Cycle})
 \begin{eqnarray}
 \label{C3} 
&& \nu = \{ (x_1,x_2), (x_2,x_3), \ldots,  (x_k, x_1)\} \\
&& \to \hat{\nu} \equiv \{ (x_1,n+x_2), (x_2,n+x_3), \ldots,  (x_k, n+x_1)\}. \nonumber
 \end{eqnarray}
 Now, for all $k\ge 3$, the  transposition  of  $x_j$ with  $n+x_j$  in a $k$-cycle $\hat{\nu}$  results in a different possible  matching  $\alphb(C_1,\ldots, C_n)$ for all $j=1,\ldots,k$,     moreover, such  transpositions are independent from each other. In the first  special case of $k=1$  the only  transposition is within a single pair   and  obviously has no effect. In the second special case of  $k=2$, e.g., $\hat{\nu} = \{(x_1,n+x_2),(x_2,n+x_1)\}$,    only one such  transposition  (of the two possible)   is independent (as we can permute the order of pairs). Summarizing,  we get a factor 
\be
F_1\equiv 2^{C_2} 2^{\sum_{k=3}^n kC_k}  
\en{C4} 
 of how many different matchings in $\Mc_{2n}$ , i.e.,  satisfying Eq. (\ref{Vind}), there are for a given  cycle decomposition of a matching  $\{\nu_1,\ldots, \nu_q\} \in (C_1,\ldots,C_n)$. What is left is to find out how many cycle decompositions  $\{\nu_1,\ldots, \nu_q\} \in (C_1,\ldots,C_n)$  there are. To this goal, let us  drop  in each cycle $\nu_s$    the second  element in each  pair, obtaining on this way a cycle decomposition of a permutation $\sigma\in S_n$ (where  the sequence of  elements, one from each pair,  by their order defines a cycle in $S_n$).  We get  a map from the cycles over  permutations belonging to $\Mc_{2n}$, acting on  the double-set, to those of the symmetric group $S_n$.    Note that  a $k$-cycle on the double-set  has a cyclic order of  the pairs:  we can choose $x_j$ with which  a cycle in Eq. (\ref{C3}) will start (recall that, by the definition of such a  cycle, we are allowed to permute  the two elements inside each pair). On the other hand, the   respective $k$-cycle of the symmetric group $S_n$ obtained by our  map, e.g.,  from the cycle of Eq. (\ref{C3}) we get  $x_1\to x_2\to \ldots \to x_k \to x_1$,  has a well-defined  specific direction, not a cyclic order.  We have that for $k\ge 3$  a $k$-cycle and its inverse from  $S_n$, while being   different cycles, nevertheless   correspond to one and the same $k$-cycle on the double-set. Thus, we have to account for the double counting of cycles from  $\Mc_{2n}$ by those from  $S_n$ by introducing the factor
 \be
F_2 \equiv  \frac{1}{\prod_{k=3}^n 2^{C_k}}
 \en{C5}
to the  number of cycles from  $S_n$  of a given   type $(C_1,\ldots,C_n)$, i.e.,  $ \#_{S_n}(C_1,\ldots,C_n)$ of Eq. (\ref{C1}). Combining the two factors $F_1$ and $F_2$ given by  Eqs.~(\ref{C4})-(\ref{C5}), while  using that  $\sum_{k=1}^nkC_k = n$,  and applying the result to $\#_{S_n}(C_1,\ldots,C_n)$ of Eq. (\ref{C1}) we obtain the total number of matchings $\alphb\in \Mc_{2n}$ corresponding to a given cycle structure 
$ (C_1,\ldots,C_n)$:
  \begin{eqnarray}
   \label{C6}
&& \#_{\Mc_{2n}}(C_1,\ldots,C_n) = \#_{S_n}(C_1,\ldots,C_n) F_1F_2  \nonumber\\
&& =   \frac{2^n n!}{\prod_{k=1}^nC_k! (2k)^{C_k}}.
\end{eqnarray}
Comparing Eq. (\ref{C1}) with Eqs.  (\ref{C2}) and (\ref{C6}), we see that  the cycle indices of  $\Mc_{2n}$ and $S_n$ are intimately related:
\be
\Xi_n(t_1,\ldots,t_n) = 2^n \Theta_n\left(t_1/2,\ldots, t_n/2\right). 
\en{C7}

 %%%%%%%%%%%%%%%%%%%%%%%%%%%%%%%%%%%%%%% 
  \section{ Probability that $n$ pairs of   photons  are   indistinguishable    } 
\label{appD}
  
Using in      Eq. (\ref{E25}) of section \ref{sec3B}   the  cycle index in  Eqs. (\ref{C1}), (\ref{C2})  and  (\ref{C7}), by  setting $t_k = \frac12\sum_j p^k_j$  we obtain:
 \begin{eqnarray}
 \label{D1}
 q^{(2n)}   &= &  \frac{1}{(2n-1)!!}\sum_{\alphb\in\Mc_{2n}} J(\alphb) \nonumber\\
 &= &  \frac{1}{(2n-1)!!}\sum_{C_1,\ldots, C_n}   \frac{2^n n!}{\prod_{k=1}^nC_k! (2k)^{C_k}} \prod_{k=2}^n(2t_k)^{C_k}\nonumber\\
& = & \frac{2^n}{(2n-1)!!}\Theta_n\left(t_1,t_2,\ldots, t_n\right).
\end{eqnarray}  
 To compute the cycle index  in Eq. (\ref{D1})  we can use the generating function approach \cite{Stanley}, which reads
 \be
 \Theta_n(t_1,\ldots, t_n) = \left(\frac{\partial}{\partial x}\right)^n \left. \exp\left\{ \sum_{k=1}^\infty t_k\frac{x^k}{k}\right\}\right|_{x=0}.
 \en{D2} 
From Eqs. (\ref{D1})-(\ref{D2}) we get 
 \begin{eqnarray} 
 && \Theta_n(t_1,\ldots,t_n) =  \left(\frac{\partial}{\partial x}\right)^n \left. \exp\left\{ \frac12 \sum_{j=1}^\infty \sum_{k=1}^\infty  \frac{(p_jx)^k}{k}\right\}\right|_{x=0} \nonumber\\
 &&=  \left(\frac{\partial}{\partial x}\right)^n   \prod_{j=1}^\infty (1-p_j x)^{-\frac12}  \Bigr|_{x=0}\nonumber\\
 && = \sum_{|\bs|=n}\binom{n}{\bs} \prod_{j=1}^\infty\left(\frac{\partial}{\partial x}\right)^{n_j}(1-p_j x)^{-\frac12}  \Bigr|_{x=0}\nonumber\\
 &&=  \sum_{|\bs|=n}\binom{n}{\bs}\prod_{j=1}^\infty \frac12 \left( \frac12  + 1\right)\ldots \left( \frac12 + s_j-1\right)p_j^{s_j}\nonumber\\
 &&= 2^{-n}  \sum_{|\bs|=n} \binom{n}{\bs} \prod_{j=1}^\infty (2s_j-1)!! p_j^{s_j},
  \end{eqnarray}
 where we  have used the Leibniz rule for derivative,    the identity $\ln(1-z) = -\sum\limits_{k=1}^\infty \frac{x^k}{k}$,  denoted by $\bs = (s_1,s_2,\ldots)$, $|\bs|\equiv  s_1+s_2 +\ldots = n$, the distribution of $n$ photons over  the internal modes $j=1,2,\ldots$ and by $\binom{n}{\bs}$ the corresponding multinomial.  The probability of Eq. (\ref{D1}) becomes
 \begin{eqnarray}
 \label{D3}
q^{(2n)} & = & \frac{1}{(2n-1)!!}  \sum_{|\bs|=n} \binom{n}{\bs}  \prod_{j=1}^\infty (2s_j-1)!!  p_j^{s_j}\nonumber\\
 &=& \binom{2n}{n}^{-1}\sum_{|\bs|=n}  \prod_{j=1}^\infty \binom{2s_j}{s_j} p_j^{s_j}, 
 \end{eqnarray} 
 where the second form is obtained with the help of the identity
 \be
 \frac{(2m-1)!!}{ m!} = 2^{-m}\binom{2m}{m}. 
 \en{D4}
 \medskip
 
 %%%%%%%%%%%%%%%%%%%%%%%%%%%%%%%%%%%%%%% 
  \section{ Probability to detect $2n$    photons     } 
\label{appE}

Let us derive the probability $  \mathring{p}(2n)$ that exactly $2n$ photons are detected at a multiport output in the ideal case. Consider  first the single-mode squeezed state $|r_k)$ as in   Eq. (\ref{Squeez1})   (with  $p_1=1$). The probability  to detect   $2n$ photons is given by projection on  the  Fock state of $2n$ photons  and reads    
\be
  \mathring{p}_k(2n) =    Z^2_k \binom{2n}{n}\left(\frac{r_k}{2}\right)^{2n},
\en{AE1}
with $Z_k =  (1-r_k^2)^{\frac14}$.  
Since the probability  to detect  a given number of     photons in all possible output configurations  is independent of the     interferometer, for a tensor product of $N$ single-mode squeezed states with the squeezing  parameters $r_1,\ldots, r_N$, the  probability to detect $2n$    photons reads 
\begin{eqnarray}
\label{AE2}
 && \mathring{p}(2n) = \sum_{|\nb|=n}\prod_{k=1}^N  \mathring{p}_k(2n_k) =   \mathring{\Zc}^2\sum_{|\nb|=n} \prod_{k=1}^N\binom{2n_k}{n_k}\left(\frac{r_k}{2}\right)^{2n_k}\nonumber\\
 \end{eqnarray}
with $ \mathring{\Zc} = \prod_{k=1}^N(1-r^2_k)^{\frac14}$.  The  expression can be simplified for the coinciding  squeezing parameters $r_k = r$. The following identity can be used 
 \begin{eqnarray}
 \label{AE3}
&&\sum_{n=0}^\infty x^n \sum_{|\nb|=n} \prod_{k=1}^N\binom{2n_k}{n_k}  = \prod_{k=1}^N\sum_{n_k=0}^\infty \binom{2n_k}{n_k} x^{n_k} \nonumber\\
&& = (1- 4x)^{-\frac{N}{2}},
\end{eqnarray}
since $\sum_{n=0}^\infty \binom{2n}{n} x^{n} = (1-4x)^{-\frac12}$. Taking the $n$-th term from the  Taylor series  of the expression  in  Eq. (\ref{AE3}) we get
\be
 \sum_{|\nb|=n} \prod_{k=1}^N\binom{2n_k}{n_k} = \frac{4^{n}}{n!} \left(\frac{N}{2}\right)_n,
\en{AE4}
where $(m)_n \equiv m(m+1)\ldots (m+n-1)$.  Using Eq. (\ref{AE4}) into Eq. (\ref{AE2}) for $r_k=r$ we obtain
\be
 \mathring{p}(2n)  =  (1-r^2)^{\frac{N}{2}} \left(\frac{N}{2}\right)_n \frac{r^{2n}}{n!}. 
\en{AE5}

%%%%%%%%%%%%%%%%%%%%%%%%%%%%%%%%%%%%%%%%%%%%%%%%%

\end{document}